\documentclass[final,12pt]{elsarticle}

\usepackage{graphicx}
\usepackage{amssymb}
\usepackage{amsmath}
\usepackage{amsthm}

\usepackage[dvipsnames]{xcolor}

\usepackage[all]{nowidow}

\biboptions{sort&compress}

\newcommand{\lambdabar}{{\mkern0.75mu\mathchar '26\mkern -9.75mu\lambda}}

\journal{Annals of Physics}

\begin{document}

\begin{frontmatter}


\title{Radiation reaction of classical hyperbolic oscillator: experimental signatures}



\author{Yuan Shi}
\ead{shi9@llnl.gov}

\address{Lawrence Livermore National Laboratory, Livermore, California 94551, USA }

\begin{abstract}
When accelerated by a constant force in the lab frame, a classical charge experiences no self force. In this case, the particle radiates without dissipating its kinetic and potential energy. But what happens when the particle enters another region with equal and opposite acceleration? Does the oscillating charge lose its mechanical energy similar to a radiating dipole, even though it seems to lose no mechanical energy within each region of constant acceleration? 
In this paper, I will show how mechanical energy is transferred to radiation energy via the Schott energy when the particle crosses the boundary between the two regions. In particular, I will show how preacceleration, which is usually regarded as an unphysical effect of the 
Lorentz-Abraham-Dirac self force, is essential for the energy transfer. Moreover, I will show that the commonly adopted Landau-Lifshitz approximation, which removes preacceleration, introduces second-order secular energy error. 
On a more fundamental level, the validity of classical electrodynamics is in fact questionable because quantum effects are likely important. The classical prediction can be tested experimentally by observing frequency chirping of radiation, whereby micro physics leaves signatures on macroscopic scales. The required experimental accuracy is estimated. Trap experiment of this type is complementary to collider experiments that endeavor to observe radiation reaction for elementary particles.
\end{abstract}


\end{frontmatter}


\section{Introduction}
\label{Sec:Intro}

Radiation reaction, the phenomenon whereby a charged particle looses energy and momentum due to its own radiation, has gained renewed interest in recent years after more than a century of purely theoretical investigations \cite{Rohrlich2000self,Mcdonald2018history}. The hope is that this historically confusing phenomenon may finally become observable for elementary particles in experiments \cite{Cole2018experimental,Poder2018experimental,Wistisen2018experimental}. 
Although radiation reaction is likely a quantum phenomena \cite{Neitz13,Blackburn14,Dinu16}, for which classical electrodynamics fails, it is still interesting to see exactly how classical physics breaks down and whether there exist regimes where the classical description remains valid.

While recent work on radiation reaction focuses on laser-plasma interactions \cite{Kumar13,Ji14,Liseykina2016inverse} and collider-type experiments \cite{Ritus1985,Bula1996observation,Hartemann1996classical,Keitel1998radiative,Piazza2009strong,Piazza10,Kravets13,Li14}, in which energetic electrons interact with intense laser pulses through nonlinear Compton scattering, here I will consider a complementary experiment where the charged particle is confined in an electrostatic trap. In trap-type experiments, the electron is accelerated by 
static electric field instead of real photons in laser fields.
In view that radiation reaction is usually an extremely tiny effect, collider-type experiments benefit from large acceleration that enhances radiation, while trap-type experiments benefit from long confinement time that allows for the accumulation of small signals.

In this paper, I will consider what classical electrodynamics can say about a charged particle oscillating in an electrostatic trap. In particular, I will consider a trap in which the restoring force has constant magnitude. 
Simple hyperbolic trap of this kind not only exhibits distinct experimental signatures, but also exposes fundamental physics that may be concealed in more messy experiments.

Constant acceleration, while seemingly simple, turns out to be a particularly tricky case for classical radiation reaction \cite{Fulton1960classical}. 
In classical electrodynamics, it is clear that charged particles radiate electromagnetic waves when accelerated with respect to the observer \cite{Rohrlich1961definition,Boulware1980radiation}, because changes in the near field propagate to the far field in the form of waves given that the speed of light is finite.
It is also clear that waves in the far field carry away energy and momentum \cite{Schild1960radiation}, so the particle should experience a finite self force in the near field, which convert the energy that is localized in the vicinity of the particle to the energy that freely propagates.

Given these two physical pictures that are intuitively correct, it was thereof very puzzling that a charged particle under constant acceleration in the lab frame does not loss any mechanical energy, which includes both the kinetic and the potential energy. 
This puzzling phenomenon can be predicted from two distinct considerations. The first consideration is the equivalence principle, which requires that all bodies fall in exactly the same way under constant gravity. Since a charged particle cannot fall behind, or ahead of, a neutral particle, it cannot exchange energy with its electromagnetic fields.
The second consideration is the self force, which can be calculated either from near-field or far-field considerations. Both calculations yield the same result, which is now known as the Lorentz-Abraham-Dirac (LAD) self force \cite{Lorentz1892theorie,Abraham1905theorie,Dirac1938classical}. The LAD self force is identically zero when the acceleration is constant in the lab frame.

Historically, it was very confusing that a radiating charged particle can somehow not feel radiation reaction. Where does the radiation energy come from? To resolve this puzzle, some of the best physicists even suggested falsely that a charged particle under constant acceleration cannot radiate at all. 
Many other physicists seek to modify the LAD self force, whose derivation involves integration by part that makes the form of the self force not unique. However, the LAD self force, which can also be derived from the modern perspective of mass renormalization \cite{coleman1961classical} and recovered as the classical limit of quantum radiation reaction \cite{Krivitskii91,Ilderton13PLB,Ilderton13PRD}, is an exact and the simplest self force for a point charge. 
Moreover, the LAD self force is the only self force for which each term involving the state of the particle has been unambiguously identified with the state of surrounding electromagnetic fields \cite{Eriksen00}.

After almost a century of confusion, it is now understood where the radiation energy comes from in the LAD framework when the particle is under constant acceleration. The key insight, first brought to light by Schott \cite{schott1912electromagnetic}, is that the self force contains two distinct terms, which correspond to two separately conserved parts of the electromagnetic stress-energy tensor \cite{Teitelboim70}. One terms is directly related to radiation in the far field and is sometimes called the radiation force, which gives rise to the radiation energy and momentum. The other term, which is required to construct a relativistic 4-force, is related to both the far field and the near field, and is sometimes called the acceleration force that gives rise to the Schott energy and momentum. 
Just like a particle carries kinetic energy due to its velocity, the Schott energy is an extra energy only a charged particle carries due to its acceleration. 
It is this Schott energy, the field energy that is bound to the particle, that provides the freely-propagating radiation energy when the charged particle undergoes constant acceleration.
In fact, the Schott energy is closely related to the self energy in quantum electrodynamics (QED). Just as the 1-loop self energy diagram is necessary in QED to remove the infrared divergence of photon emission, the Schott term is necessary in classical electrodynamics to construct the self force when the charged particle radiates.

The physical picture we now have is the following. Under constant acceleration, the Coulomb field of the charged particle was infinitely compressed on the event horizon \cite{Bondi1955field}. As the particle decelerates from the speed of light, its radiation power is a constant. The radiation energy increases linearly, while the Schott energy decreases linearly at exactly the same rate such that the sum of the radiation energy and the Schott energy is a constant \cite{Gron2011significance}. In this way, the decrease of the Schott energy completely accounts for the increase of the radiation energy, 
and the charged particle radiates without losing any of its mechanical energy. 
In other words, an uniformly accelerated charged particle serves as the ``catalyst" that converts bounded field energy into freely propagating radiation energy.

That the Schott energy can decrease to negative infinity to compensate for the radiation energy is somewhat unpleasant. It is nevertheless a valid physical picture in the thought experiment where the point charge is accelerated by a constant force indefinitely. A more physical scenario has been considered where the charged particle enters and then leaves a region of constant acceleration \cite{Ng93,Eriksen2002electrodynamics}, which can happen, for example, when the charged particle passes between two capacitor plates. It is clear that not much should happen long before the particle enters the region and long after the particle leaves the region. It is also relatively clear that if the region is sufficiently large, the particle should behave as if it has always been experiencing the constant acceleration. However, what happen during boundary crossings are not so trivial .

Technically, for a point charge, the time it takes to cross the sharp boundary is exactly zero. This is problematic for classical electrodynamics, which is expected to be valid only on time scales that are much larger than the Compton time.
Eriksen and Gr{\o}n considered two resolutions to this problem \cite{Eriksen2002electrodynamics}. In one resolution, they compute what extra force is necessary in order to enforce a zero acceleration outside the region while a constant acceleration inside the region. They find that impulses that deliver energy and momentum to the particle are necessary when the point charge crosses the sharp boundary.
In the other resolution, only the force that provides the constant background acceleration and the LAD self force are present. The LAD model, which depends on the time derivative of the acceleration, has long been known to have two intertwined unphysical effects: the existence of runaway solutions and the presence of preacceleration. When the initial conditions is chosen such that runaway is not excited, preacceleration causes smooth changes of the acceleration when the particle crosses the boundary. Since the acceleration is no longer constant there, the self force is nonzero and the charged particle loses its mechanical energy during boundary crossings.

\begin{figure}[b!]
\centering
\includegraphics[width=0.35\linewidth]{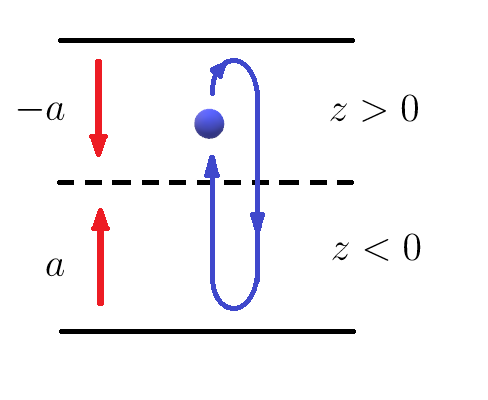}
\caption{Charged particle oscillating between two regions of equal but opposite acceleration. The constant background acceleration $a$, which can be provided by electric field between parallel capacitor plates, is in the $+z$ direction for $z<0$, and in the $-z$ direction for $z>0$.  In the simplest scenario, the point charge, which is depicted here by the blue ball, oscillates along the $z$ axis. The one-dimensional trajectory is depicted with a horizontal offset for clarity.  }
\label{fig:trap}
\end{figure}

While Eriksen and Gr{\o}n's thought experiment is a collider-type experiment, where the particle enters and leaves the system with well-defined asymptotic states, the thought experiment I consider here is a trap-type experiment, where the particle is confined in the system for an extended period of time.
In this paper, I will first employ the preacceleration solution to analyze what happens when the charged particle oscillates between two regions of constant acceleration (Fig.~\ref{fig:trap}). 
Although preacceleration is usually regarded as an unphysical effect, it arises only when one takes the point-particle limit in the classical theory \cite{Yaghjian2010relativistic}. The classical theory is free from pathologies if the charged particle is treated as an extended body to allow sufficient time for internal stress to propagate 
\cite{Rohrlich1997dynamics,Smorenburg2014classical}. The internal dynamics of the extended body may be nontrivial \cite{Bhabha1941general,Rowe75,Honig1983general,Lozada1989general,Aguirregabiria06,Medina2006radiation} and quantum effects are likely important \cite{Moniz1977radiation,Johnson2002stochastic}. However, the detailed internal dynamics is beyond the scope of classical electrodynamics. The concern of this paper is to predict refutable experimental signatures using the simplest but otherwise exact classical radiation reaction model.

Another resolution uses perturbative methods to remove unphysical effects of LAD. The perturbative approach, first adopted by Eliezer \cite{Eliezer1948classical} and independently by Landau and Lifshitz \cite{Landau1971classical}, is based on the argument that the self force is usually subdominant than the external force. One can then substitute in lower-order force laws into the LAD formula to eliminate the time derivative of acceleration. The resultant equations only contain first-order velocity derivative, and the dynamics is restricted to the critical manifold on which runaway is not excited and preacceleration is not explicitly introduced \cite{Spohn2000critical}. 
Many other variations of the perturbative approach have been advocated in the literature \cite{Mo71,Ford1991radiation,Rohrlich2002dynamics}. However, unless the infinite series is summed up,  perturbative solutions have finite errors. 
While these errors may be small and transient in collider-type experiment, they can be accumulated in trap-type experiment and affect long-time behavior of the oscillator.
As I will show in this paper, the lowest-order Landau-Lifshitz (LL) type reduction of order will introduce second-order secular energy error. The growing error may be hiding for a harmonic oscillator, but they become clearly exposed for a hyperbolic oscillator. 
Therefore, the seemingly simple case that a charged particle oscillates between two regions of constant acceleration is in fact a challenging case for classical electrodynamics.

The goal of this paper is to analyze classical predictions for the trapped particle, which would undergo simple hyperbolic oscillation in the absence of radiation reaction.
This paper is organized as follows. 
In Sec.~\ref{sec:trajectory}, I will solve the classical equation of motion that includes the LAD self force. 
In Sec.~\ref{sec:energy}, I will discuss how energy of the trapped particle is distributed among different components. 
The short-time dynamics will be further analyzed in Sec.~\ref{sec:time}, where I will show how it leads to long-time behavior using an envelope approximation. 
In Sec.~\ref{sec:chirp}, I will present frequency chirping as a distinct experimental signature of radiation reaction in the hyperbolic trap. 
In Sec.~\ref{sec:diff}, I will estimate the required experimental accuracy needed for excluding radiation reaction models based on discrepancies between the LAD and the LL self force.
The discrepancy is largest near the cutoff time, where quantization becomes important as I will show in Sec.~\ref{sec:quantization}.
Discussions are given in Sec.~\ref{sec:conclusion} followed by a summary.


\section{Classical trajectory of the oscillating charge \label{sec:trajectory}}
\subsection{Equation of motion}

For a point charge, its trajectory is described by the relativistic Newton's equation $\dot{p}^\mu=f^\mu_{\text{ext}}+f^{\mu}_{\text{self}}$, where $p^\mu=mu^\mu$ is the 4-momentum, the dot denotes derivative with respect to the proper time, and $f^\mu_{\text{ext}}$ is the external 4-force. 
In the LAD framework, the self force $f^\mu_{\text{self}}$ is given by
\begin{equation}
    \label{eq:LAD}
    f^\mu_{\text{LAD}}=\frac{2\alpha\hbar}{3c^2}\Big(u^\mu\frac{\dot{u}^{\nu}\dot{u}_{\nu}}{c^2}+\ddot{u}^\mu\Big),
\end{equation}
where $\alpha$ is the fine structure constant, $\hbar$ is the Planck constant, and $c$ is the vacuum speed of light. The first term is the radiation force $-u^\mu P_{\text{rad}}/c^2$, where $P_{\text{rad}}=-2\alpha\hbar\dot{u}^\nu\dot{u}_\nu/3c^2>0$ is the Lorentz-invariant radiation power given by the relativistic Larmor formula.
The second term is the acceleration force, which is sometimes also called the Schott term. This term is necessary to construct a 4-force that obeys $u_\mu f^\mu=0$, which is required in order for the Lorentz scalar $u_\mu p^\mu=mc^2$ to remain a constant of motion.

To solve the relativistic equation of motion, it is convenient to use hyperbolic coordinate. 
Consider one dimensional motion along the $z$ axis, then the 4-velocity $u^{\mu}/c=\gamma(1,0,0,\beta)$, where $\beta=v_z/c$ and $\gamma=1/\sqrt{1-\beta^2}$. We can introduce rapidity $w$, such that
\begin{equation}
    \beta=\tanh w.
\end{equation}
Then, the Lorentz factor becomes
\begin{equation}
    \gamma=\cosh w,
\end{equation}
and the 4-velocity becomes $u^{\mu}/c=(\cosh w, \sinh w)$. Here, I have omitted the $x$ and $y$ components for simplicity, which are always zero for one dimensional motion along the $z$ axis. Taking derivative with respect to the proper time, the 4-acceleration is $\dot{u}^\mu/c=\dot{w}(\sinh w, \cosh w)$. The term that appears in the Larmor formula is then 
$\dot{u}^\nu\dot{u}_\nu/c^2=-\dot{w}^2$, and the proper time derivative of the acceleration is $\ddot{u}/c=\ddot{w}(\sinh w, \cosh w)  +\dot{w}^2u^\mu/c$.
Notice that the last term of the acceleration force exactly cancels the radiation force, so the LAD self force can be written as 
\begin{equation}
    f^\mu_{\text{LAD}}=\frac{2\alpha\hbar}{3c}\ddot{w}(\sinh w, \cosh w),
\end{equation}
where I have again suppressed the $x$ and $y$ components. Notice that when the external force is a constant in the lab frame, $\dot{w}$ is a constant so \mbox{$\ddot{w}=0$}. In other words, the LAD self force is zero when the particle is under constant acceleration.
Finally, the external 4-force is constructed from the lab-frame 3-force $\mathbf{F}_{\text{ext}}$ as $f^\mu_{\text{ext}}=\gamma(\boldsymbol{\beta}\cdot\mathbf{F}_{\text{ext}},\mathbf{F}_{\text{ext}})$, so that $u_\mu f^\mu=0$ is satisfied. Now that the motion is one dimensional with the external force along the $z$ direction, we can write $f^\mu_{\text{ext}}=F_{\text{ext}}(\sinh w, \cosh w)$. 
Since all terms in the relativistic Newton's equation are proportional to the vector $(\sinh w, \cosh w)$, the Newton's equation is satisfied if and only if
\begin{equation}
\label{eq:EOM}
    \dot{w}=\frac{a_{\text{ext}}}{c}+\tau_0\ddot{w},
\end{equation}
where $a_{\text{ext}}=F_{\text{ext}}/m$ is the acceleration in the lab frame. The characteristic time scale of the self force is 
\begin{equation}
    \label{eq:tau0}
    \tau_0=\frac{2\alpha\hbar}{3mc^2},
\end{equation}
which is roughly the time it takes for light to cross the classical electron radius and
is only a small fraction of the Compton time. 
It is alarming that the presumably classical self force turns out to act on sub-Compton time scale, where classical electrodynamics is not expected to be valid.

\subsection{Initial value problem}
Within the framework of classical electrodynamics, the goal now is to solve the classical equation of motion. Let us normalize the proper time $\tau$ such that $\tau=\tau_0 T$, and introduce the normalized acceleration $A=a_\text{ext}\tau_0/c$. Then, the equation we need to solve is simply
\begin{equation}
\label{eq:nEOM}
    \dot{w}=A+\ddot{w},
\end{equation}
where dot now denotes derivative with respect to the normalized proper time. The 
general solution of the above equation is
\begin{equation}
    \dot{w}=e^{T-T_0}\Big[\dot{w}_0-\int_{T_0}^T e^{-(T'-T_0)}A(T') dT'\Big],
\end{equation}
where the constant $\dot{w}_0$ is the initial value of $\dot{w}$ at $T=T_0$. Since the rapidity $w$ is the equivalence of velocity in the hyperbolic coordinate, the time derivative $\dot{w}$ is the equivalence of acceleration. 

In the usual Newtonian mechanics, only initial position and velocity need to be specified. Here, the initial value of acceleration is also required due to the LAD self force. This additional degree of freedom allows a new class of solutions, which exhibit unphysical runaway behavior. For runaway solutions, the acceleration grows even in the absence of the external force. The choice of $\dot{w}_0$ that does not excite the runaway mode is given by \cite{Rohrlich1961equations}
\begin{equation}
    \dot{w}_0=\int_{T_0}^{+\infty}e^{-(T'-T_0)}A(T')dT'.
\end{equation}
With this choice of the initial acceleration, the solution to the equation of motion 
can be written as
\begin{equation}
    \label{eq:pre}
    \dot{w}_{\text{PA}}(T)=e^T\int_T^{+\infty} e^{-T'}A(T') dT'.
\end{equation}
In the special case where $A$ is a constant, $\dot{w}_{\text{PA}}=A$ is also a constant. Then, $\ddot{w}_{\text{PA}}=0$ is consistent with the self force being zero under constant acceleration.
Unfortunately, in more general cases, the above choice of $w_0$ that eliminates runaway excites preacceleration, whereby the acceleration at the present time is influenced by the force at future time. Preacceleration is usually regarded as another unphysical effect of the LAD self force, because it violates classical causality.

To make further progress without invoking quantum mechanics, we have to accept pathologies of classical electrodynamics. In this section, I choose to trade runaway for preacceleration, and accept Eq.~(\ref{eq:pre}) as a valid solution of the equation of motion.
For a charged particle trapped between two regions of equal but opposite accelerations, the normalized acceleration
\begin{equation}
    A=-\frac{a\tau_0}{c}\text{sgn}(z),
\end{equation}
where $a$ is the absolute value of the constant accelerations in the lab frame.
Notice that in Eq.~(\ref{eq:pre}), we need $A$ as a function of $T$ instead of a function of $z$. Suppose the particle starts from $z_0=0^+$, right above the midplane, with initially positive rapidity $w_0>0$ at time $T_0$. Furthermore, suppose the particle subsequently returns to the midplane at time $T_1, T_2\dots$. Then, for $T_{n-1}<T<T_{n}$, the normalized acceleration
\begin{equation}
    \label{eq:An}
    A=\epsilon(-1)^n,
\end{equation}
where $\epsilon=\tau_0 a/c$ is typically a very small number in order for classical electrtodynamics to hold. To get a sense of how small $\epsilon$ is, we can insert electron mass into Eq.~(\ref{eq:tau0}), then the time constant \mbox{$\tau_0\approx 6.3\times10^{-24}$ s}. Suppose the acceleration is provided by an electric field $E=1$ MV/m, then $a\approx 1.8\times 10^{17} \, \text{m/s}^2$. Even with such a large acceleration, the normalized acceleration $\epsilon\approx 3.7\times 10^{-15}$ is still a very tiny number.
In fact, when the acceleration is provided by the electric field, another way to see why $\epsilon$ must be a very small number is to note $\epsilon=2\alpha E/3 E_c$, where $E_c=m^2c^3/e\hbar\approx$\mbox{$1.3\times 10^{18}$ V/m} is the critical field of QED. In order for classical electrodynamics to hold, we need $E\ll E_c$. The normalized acceleration $\epsilon$ is thus an even smaller number than $E/E_c$ when multiplied by the fine structure constant $\alpha\approx 1/137$. We see $\epsilon\ll 10^{-2}$ within the realm of classical electrodynamics.

Given the normalized external acceleration as a function of time, we can then determine the rapidity of the trapped particle.
Substituting Eq.~(\ref{eq:An}) into Eq.~(\ref{eq:pre}), for $T_{n-1}<T<T_{n}$, we have
\begin{eqnarray}
    \label{eq:dwT}
    \nonumber
    \dot{w}_{\text{PA}}(T)&=&e^T\Big(\int_T^{T_n}+ \int_{T_n}^{T_{n+1}}+\dots\Big) e^{-T'} A(T') dT'\\
    &=&\epsilon(-1)^n\Big[1-2\sum_{k=0}^{\infty}(-1)^k e^{T-T_{n+k}}\Big].
\end{eqnarray}
It is easy to see that $\dot{w}_{\text{PA}}(T)$ is a continuous function, so its integral $w_{\text{PA}}(T)$ is also continuous. Denoting $w_{n-1}$ the value of $w$ at $T=T_{n-1}$, then for $T_{n-1}<T<T_{n}$, we can integrate $\dot{w}_{\text{PA}}(T)$ to find the rapidity
\begin{equation}
    \label{eq:wT}
    w_{\text{PA}}(T)=w_{n-1}+\epsilon(-1)^n\Big[\big(T-T_{n-1}\big)-2\big(e^{T}-e^{T_{n-1}}\big)\sum_{k=0}^{\infty}(-1)^k e^{-T_{n+k}}\Big].
\end{equation}
The velocity can be further integrated to find the trajectory of the particle, and the initial value problem would be solved.

Unfortunately, we do not know the zero-crossing time. Without knowing $T_n$, the above solution is only symbolic. To find the zero-crossing time, let us consider what happens for $T_{n-1}<T<T_{n}$. During this time, the particle is first decelerated until it comes to a complete stop and then accelerated in the opposite direction and returns to the midplane. In other words, the displacement is zero when the particle starts from $z=0$ at $T_{n-1}$ and then returns to $z=0$ at $T_n$. 
Since the lab time is related to the proper time by $dt/d\tau=\gamma$, we have the following constraint for zero displacement:
\begin{equation}
    \label{eq:constraint}
    0=\int_{T_{n-1}}^{T_n} \beta \gamma \,dT=\int_{T_{n-1}}^{T_n} \sinh w(T)\, dT.
\end{equation}
In principle, if we can solve the above constraint for $n=1,2,\dots$, then all zero-crossing time can be determined, from which we can obtain the trajectory of the trapped particle. However, due to preacceleration, within each time interval, we always need the complete information all the way up to the infinite future, which we have no knowledge of in an initial value problem.

\subsection{Final value problem}
Although information from the future is suppressed exponentially, which allows for truncation of the infinite series, here I will take a different approach to approximate the solution. Suppose we watch the charged particle and turn off the electric field after the particle has crossed the midplane for exact $N$ times. Then, after the external force is turned off, the particle will free stream with its final velocity. In this way, what happens in the infinite future becomes known, and instead of solving the initial value problem, we can solve the final value problem where the information propagates backward in time. We can then take the limit $N\rightarrow\infty$ to determine what happens when the charged particle is trapped indefinitely.

Starting from the final time, since the external acceleration is zero and the runaway mode is not excited, we have $\dot{w}(T)=0$ for $T>T_N$.
At earlier time when $T_{n-1}<T<T_{n}$, where $n\le N$, the infinite sum in Eq.~(\ref{eq:dwT}) terminates, and we have
\begin{equation}
    \label{eq:NdwT}
    \dot{w}_N(T)=\epsilon \big[(-1)^n -e^T S_{N,n}\big],
\end{equation}
where the series $S_{N,n}=2\sum_{k=n}^{N}(-1)^k e^{-T_{k}}-(-1)^N e^{-T_N}$.
The rapidity at earlier time can be found by integrating Eq.~(\ref{eq:NdwT}) backward in time once $w_{N,n}:=w_N(T=T_n)$ is known.
Now that the future information is known, the prior zero-crossing time $T_{n-1}$ can be solved. 
Denoting $d_{n-1/2}=T_n-T_{n-1}$, the integral constraint Eq.~(\ref{eq:constraint}) can be approximated by
\begin{eqnarray}
    \label{eq:constraint_app}
    0\simeq\int_0^{d_{n-1/2}}dT\,\sinh (\epsilon T-\nu)=\frac{2}{\epsilon}\sinh\Big(\frac{\epsilon d_{n-1/2}}{2}\Big) \sinh\Big(\frac{\epsilon d_{n-1/2}}{2}-\nu\Big),
\end{eqnarray}
where $\nu=(-1)^n(w_{N,n}+\epsilon S_{N,n} e^{T_n})$. Here, I have made the approximation that $(1-e^{-T})\simeq 1$, which is valid when $T\gg 1$. Since classical electrodynamics requires that the oscillation half period $d_{n-1/2}\gg1$, the integral is well approximated for $1\ll T<d_{n-1/2}$. 
The oscillation half period can then be estimated by 
\begin{equation}
    \label{eq:dn}
    d_{n-1/2}\simeq 2(-1)^n \Big(\frac{w_n}{\epsilon}+R_{N,n}\Big)\simeq \frac{2|w_{n-1}|}{\epsilon}.
\end{equation}
Here, $(-1)^n w_n=|w_n|$ because of the particle is initially heading in the $+z$ direction. To leading order, $R_{N,n}:=e^{T_n} S_{N,n}$ can usually be dropped in comparison with $w_n/\epsilon$.
The last approximation will become clear later [Eq.~(\ref{eq:adw})] when we retain the next-to-leading order corrections.

The procedure for solving the final value problem is as follows. First, given the final zero-crossing time $T_N$, we know the final value $R_{N,N}=(-1)^N$.
Second, given the final value $w_{N,N}$, the rapidity when $T_{n-1}<T<T_n$ can be computed 
by integrating backward in time:
\begin{equation}
    \label{eq:NwT_prac}
    w_N(T)=w_{N,n}+\epsilon\big[(-1)^n\big(T-T_{n}\big)+\big(1-e^{T-T_n}\big)R_{N,n}\big].
\end{equation}
Third, we can solve for the oscillation half period $d_{n-1/2}$ such that the integral constraint is satisfied:
\begin{equation}
    0=\int_0^{d_{n-1/2}} dT\, \sinh w_N(T_n-T) .
\end{equation}
Here, we need to solve a transcendental equation. This can be done numerically using, for example, Newton's method, with the initial guess given by Eq.~(\ref{eq:dn}).
Having determined the value of $d_{n-1/2}$, we can then compute $R_{N,n-1}$ from the backward recurrence relation
\begin{equation}
    \label{eq:Rnn}
    R_{N,n-1}=2(-1)^{n-1}+e^{-d_{n-1/2}} R_{N,n},
\end{equation}
and compute $w_{n-1}$ by substituting $T_{n-1}=T_n-d_{n-1/2}$ into Eq.~(\ref{eq:NwT_prac}).
These steps can be repeated backward in time for $n\le N$ until $n=0$ when we reach the initial time.

When solving the final value problem, the independent variables are $T_N$ and $w_{N,N}$, and we do not have direct control over $T_0$ and $w_0$. In order to make connection with the initial value problem, we can shift the time axis such that $T_0$ is fixed. Moreover, we can recursively adjust $w_{N,N}$ such that the backtracted $w_{N,0}$ agrees with $w_0$. 
Once we determine $w$, the displacement can be easily found by integration using the condition that $z=0$ at $T_n$.
Denoting $w_N(T)$ the rapidity trajectory with $N$ returns, the limit of the function series $w_\infty(T)=\lim_{N\rightarrow\infty}w_N(T)$ is then the trajectory when the particle is confined in the electrostatic trap indefinitely.

Due to radiative losses, the particle oscillates with decaying amplitude. As the amplitude decays, the oscillation period also decays. In its final state, the particle will loose all its mechanical energy and stays at rest on the $z=0$ plane.
Although it is a legitimate mathematical question to ask exactly how the final state is reached and how the function series $\{w_N(T)\}_{N=1}^{\infty}$ converges, these questions are of little physical significance, because long before reaching the final state, the particle already oscillates on time scales shorter than $\tau_0$, for which classical electrodynamics is likely invalid. Therefore, the primary concern here is what happens on some finite time interval $T\in[T_0, T_f]$, when the oscillation period is much larger than the Compton time.

\begin{figure}[b!]
\centering
\includegraphics[width=0.75\linewidth]{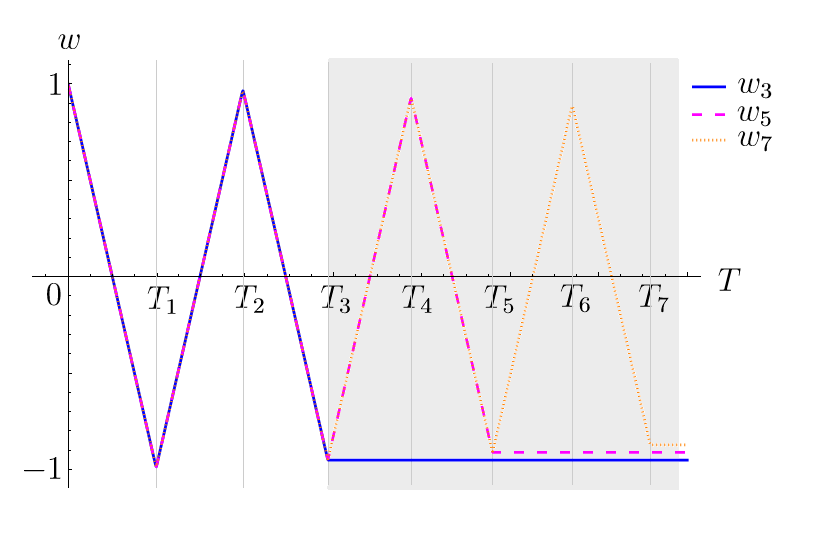}
\caption{Rapidity trajectories $w_N(T)$ for $N=3$ (solid blue), $5$ (dashed magenta), and $7$ (dotted orange) show rapid convergence to $w_\infty(T)$ for $T\in[T_0, T_3]$. All trajectories have the same initial value $w_0=1$ at $T_0=0$, and the normalized acceleration $\epsilon=0.01$. The rapidity changes almost linearly as expected for hyperbolic motion, except near $T\approx T_n$, where $|\dot{w}|$ is smoothly reduced due to preacceleration. }
\label{fig:Sequence}
\end{figure}

An example of the above numerical procedure for solving the final value problem is illustrated in Fig.~\ref{fig:Sequence}. Each trajectory is obtained by adjusting the value of $w_{N,N}$ such that the initial value $w_0=1$ is fixed. In this example, the normalized acceleration $\epsilon=0.01$ is considered large for classical electrodynamics. 
Even so, the influence of preacceleration is limitted to $T\approx T_n$, during which the absolute value of $\dot{w}$ is smoothly reduced, so that the particle is not able to return to its original rapidity after each period of oscillation. 
Away from the midplane, the particle travels with almost constant acceleration as expected for hyperbolic motion.
The oscillation amplitude slowly decreases, so does the oscillation period, until the external force is turned off after $N$ oscillations.
In this example, notice that the three trajectories are barely distinguishable on the time interval $[T_0,T_3]$. On this fixed time interval, although $w_5$ receives corrections from the future when compared to $w_3$, the corrections are exponentially suppressed. The corrections $w_7$ receives from the future are suppressed even further. Due to the exponential suppression, the function series converges rapidly for $T\in[T_0,T_3]$.
When $\epsilon$ is smaller or when $w_0$ is larger, the oscillation half period is larger, so the influence of preacceleration is comparatively smaller and the function series converges faster.
For given $\epsilon$ and $w_0$, it can be proven in a mathematically rigorous way that the function series $\{w_N(T)\}_{N=1}^{\infty}$ uniformly converges on any given time interval $[T_0, T_f]$.
In practice, if we are interested in the dynamics of $w_\infty$ for $T\in[T_0, T_N]$, then $w_{N+1}$ already provides a very good approximation.

\section{Energy budget of the trapped particle \label{sec:energy}}
From the previous example, we see that each time the particle returns to the midplane, it looses some of its rapidity due to preacceleration. This can be seen more clearly from the particle's energy budget.
The energy of the particle can be separated into four components: the kinetic energy, the potential energy, the radiation energy, and the Schott energy. 
In this section, I will investigate how these energy components evolve in details.

Among the four energy components, the kinetic energy and the potential energy can be defined for charged particles as well as neutral particles, and their sum is usually referred to as the mechanical energy.
The kinetic energy is a familiar concept in classical mechanics. It is the energy a particle possess due to its motion. In special relativity, the kinetic energy $E_K=mc^2(\gamma-1)$ is the energy of the particle in excees of its rest energy. Denoting $\mathcal{E}_K:=E_K/mc^2$, then in the hyperbolic coordinate, the normalized kinetic energy
\begin{equation}
    \label{eq:EK}
    \mathcal{E}_K=\cosh w-1.
\end{equation}
The potential energy $E_P$ is also a familiar concept, and equals to the negative work done by conservative forces. 
In differential form, $dE_P/dz=-F_{\text{ext}}$. Now that the external force is constant, $\mathcal{E}_P=\epsilon|z|/\tau_0c$,
where $\mathcal{E}_P:=E_P/mc^2$ is the normalized potential energy. Here, I have taken $z=0$ to be the reference point where $\mathcal{E}_P=0$.
Alternatively, we can express the potential energy as a function of time:
\begin{eqnarray}
    \label{eq:dEp}
    \frac{d\mathcal{E}_P}{dT}&=&\frac{d\mathcal{E}_P}{dz}\frac{dz}{dT}=-\frac{F_{\text{ext}}}{mc^2}\tau_0 c\gamma\beta=-A\sinh w.
\end{eqnarray}
When $T_{n-1}<T<T_{n}$, the above can be integrated to give
\begin{equation}
    \label{eq:Ep}
    \mathcal{E}_P(T)=\epsilon(-1)^{n-1}\int_{T_{n-1}}^{T} dT' \sinh w(T').
\end{equation}
Once we have determined the rapidity trajectory $w(T)$, both the kinetic and the potential energy can be readily evaluated using the above formulas.

The other two energy components, namely, the radiation energy and the Schott energy, are less familiar concepts, which are nonzero only for charged particles. 
The radiation energy $E_R$ is the energy carried by the freely propagating electromagnetic waves in the far field, which originates from the charged particle when it undergoes acceleration. Using the Larmor formula, the rate at which the radiation energy change is
\begin{equation}
    \label{eq:dEr}
    \frac{d\mathcal{E}_R}{dT}=\gamma  \frac{\tau_0 P_{\text{rad}}}{mc^2}=\dot{w}^2\cosh w,
\end{equation}
where $\mathcal{E}_R:=E_R/mc^2$ the normalized radiation energy. The $\gamma$ factor arises because while electromagnetic waves propagate in the lab time, the particle looses energy in its proper time.
Suppose we choose the energy reference $\mathcal{E}_R=0$ at $T_0$, then the radiation energy at later time can be computed by
\begin{equation}
    \label{eq:Er}
    \mathcal{E}_R(T)=\int_{T_0}^T dT' \dot{w}^2(T')\cosh w(T').
\end{equation}
Finally, the Schott energy is constituted of the near field energy as well as the interference between the near field and the far field \cite{Teitelboim70}. This electromagnetic energy can be expressed as a state function of the charged particle \cite{Eriksen00}, which equals to the negative work done by the acceleration force. 
The acceleration 4-force is the second term of the LAD self force: $2\alpha\hbar\ddot{u}^\mu/3c^2$. Integrating the negative 4-force in time, the Schott 4-momentum is then $p_S^{\mu}=-2\alpha\hbar\dot{u}^\mu/3c^2$. The Schott energy $E_S=cp_S^0$ is the time component of $p_S^{\mu}$. In the hyperbolic coordinate, the normalized Schott energy is
\begin{equation}
    \label{eq:Es}
    \mathcal{E}_S=-\dot{w}\sinh w.
\end{equation}
It is sometimes advocated in the literature that the Schott energy be absorbed into the mechanical energy by renormalizing the mass of the charged particle. However, keeping the Schott energy as a separate term appears to be more illuminating. In fact, the Schott energy is to the radiation energy as the kinetic energy is to the potential energy. These two energy pairs are analogous: both the Schott and the kinetic energy are given as state functions of the particle, while both the radiation and the potential energy are given in terms of integrals.

The above decomposition is meaningful because the total energy is conserved while being distributed dynamically among the four components, each with a distinct physical meaning. To show that the total energy is conserved, we can show that its time derivative is zero. The time derivative of the kinetic energy is $\dot{\mathcal{E}}_K=\dot{w}\sinh w$, and the time derivative of the Schott energy is $\dot{\mathcal{E}}_S=-\ddot{w}\sinh w-\dot{w}^2\cosh w$. Combining these with Eq.~(\ref{eq:dEp}) and (\ref{eq:dEr}), the time derivative of the normalized total energy $\mathcal{E}=\mathcal{E}_K+\mathcal{E}_P+\mathcal{E}_R+\mathcal{E}_S$ is
\begin{equation}
    \label{eq:dE}
    \frac{d\mathcal{E}}{dT}=(\dot{w}-\ddot{w}-A)\sinh w=0.
\end{equation}
A similar conservation law exists for the total momentum, with $\sinh w$ replaced by $\cosh w$ in the above expression.
We see the conservation of the total energy and momentum is guaranteed by the LAD equation of motion [Eq.~(\ref{eq:nEOM})]. Unlike a neutral particle whose conserved total energy is distributed only among the kinetic and the potential components, here for a charged particle, the energy is further distributed towards the radiation and the Schott energy. Notice that the radiation energy can come not only from the mechanical energy, but also from the Schott energy. In particular, in the special case of hyperbolic motion, where the accleration is constant, we have $\ddot{w}=0$. In this case $\dot{\mathcal{E}}_R=-\dot{\mathcal{E}}_S$, and the radiation energy comes entirely from the Schott energy. In other words, under constant acceleration, the charged particle looses no mechanical energy despite of its radiation.

\begin{figure}[t]
	\centering
	\includegraphics[width=0.95\linewidth]{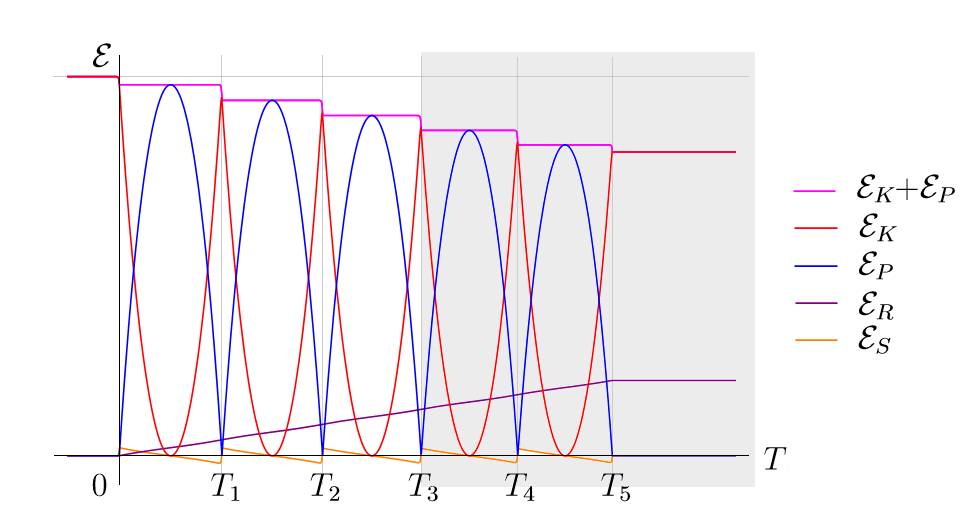}
	\caption{Time evolution of energy components for the rapidity trajectory $w_5$, where $w_0=1$ and $\epsilon=0.01$. This trajectory provides a good approximation for $w_\infty$ on the time interval $[T_0,T_3]$. Between adjacent zero crossings, the mechanical energy (magenta) is roughly constant and distributed among the kinetic (red) and the potential (blue) energy; the radiation energy (purple) increases at the expanse of the Schott energy (orange). During zero crossings $T\approx T_n$, the mechanical energy is consumed to charge the Schott energy. This process happens on $T=\tau/\tau_0\sim 1$ time scale due to preacceleration.}
	\label{fig:Energy}
\end{figure}

When the charged particle undergoes hyperbolic oscillation, the constant external force
flips its sign near the midplane and the energy budget becomes quite intriguing. In Fig.~\ref{fig:Energy}, different energy components are plotted for $w_5(T)$ in the example where $w_0=1$ and $\epsilon=0.01$. As discussed earlier, $w_5$ already provides a very good approximation for $w_\infty$ on the time interval $[T_0, T_3]$.
As can be seen from the figure, the mechanical energy (magenta) is roughly constant between adjacent zero crossings. 
The kinetic energy (red) first decreases due to deceleration, and then increases when the particle is accelerated in the opposite direction. 
Complementarily, the potential energy (blue) first increases when the particle travels against the external force, and then decreases when the particle travels along the force. 
Interesting dynamics occurs on $T\sim 1$ time scales when $T\approx T_n$, where the mechanical energy drops rapidly to charge the Schott energy (orange). The Schott energy jumps near $T\approx T_n$ because $\ddot{w}$ receives a kick from the sign flip of the external force. 
The Schott energy, which is recharged during every zero crossings, is subsequently discharged through the radiation energy (purple). To a very good approximation, the radiation energy increases linearly between zero crossings, while the Schott energy decreases linearly at the same rate to compensate for the radiative energy loss. 
The total energy (grey) remains constant.

\section{From short-time dynamics to long-time behavior \label{sec:time}}
\subsection{Short-time dynamics}
\begin{figure}[t]
\centering
\includegraphics[width=0.9\linewidth]{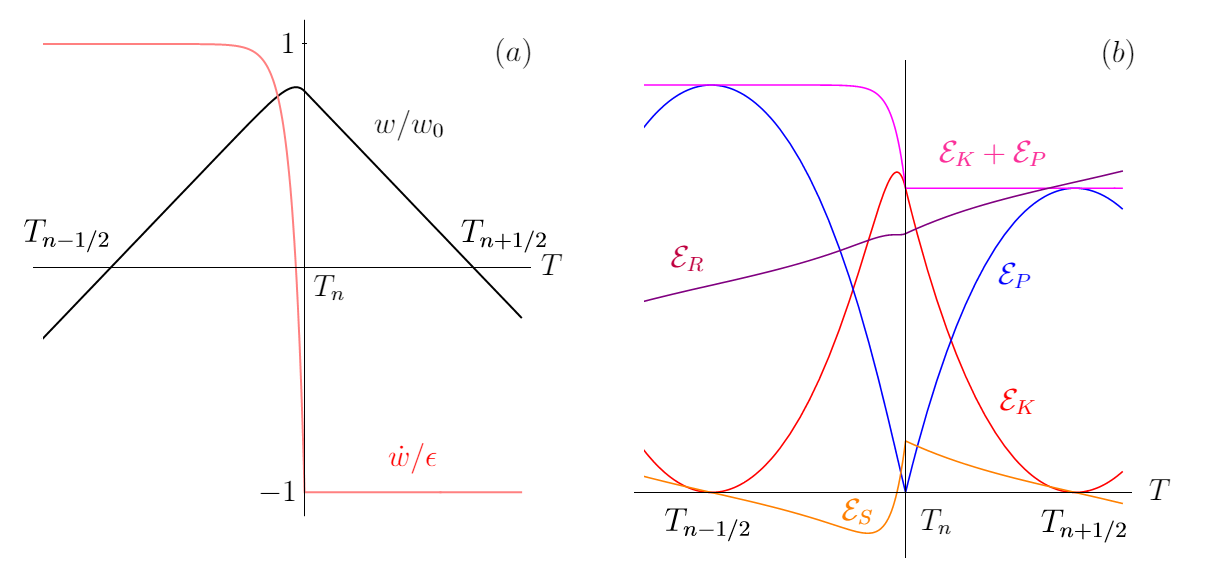}
\caption{Short-time dynamics of particle trajectory (a) and energy components (b) near $T_n$. To exaggerate deviations from hyperbolic motion, $\epsilon=0.1$ is taken excessively large. At $T_{n-1/2}$, the particle reaches maximum displacement, where the rapidity $w=0$, the kinetic energy $\mathcal{E}_K=0$, and the potential energy $\mathcal{E}_P$ reaches a maximum. At later time, the particle is accelerated towards $z=0$, during which the mechanical energy is conserved and the radiation energy $\mathcal{E}_R$ increases at the expense of the Schott energy $\mathcal{E}_S$. When $T$ increases towards $T_{n}$, the acceleration $\dot{w}$ smoothly flips sign due to preacceleration, which charges $\mathcal{E}_S$ at the expense of the mechanical energy. After crossing $z=0$, the particle is decelerated, while $\mathcal{E}_R$ continues to increase by discharging $\mathcal{E}_S$.  }
\label{fig:Short}
\end{figure}

Let us take a closer look at what happens during zero crossings, which turns out to determine the dynamics on longer time scales. We can zoom in near $T\approx T_n$, and an example is shown in Fig.~\ref{fig:Short}. In this example, $n$ is an even number, so $w_n>0$. In order to exaggerate deviations from hyperbolic motion, $\epsilon=0.1$ is taken excessively large. 
This is mathematically allowed once we are given the equation of motion.
When $\epsilon$ is larger, the acceleration is larger and the oscillation period is shorter, so that a larger portion of the trajectory is noticeably affected by preacceleration.

Across the midplane, the external force flips sign and the rapidity no longer changes linearly due to preacceleration [Fig.~\ref{fig:Short}(a)]. 
Denoting $\Delta T=T-T_n$, then for $\Delta T<0$, the acceleration [Eq.~(\ref{eq:NdwT})] can be approximated by
\begin{equation}
    \label{eq:dwn}
    \dot{w}\simeq \epsilon (-1)^n(1-2e^{\Delta T}).
\end{equation}
Here, I have used $d_{n-1/2}\gg 1$ to obtain the approximation $R_{N,n}\simeq 2(-1)^n$ from its recurrence relation [Eq.~(\ref{eq:Rnn})]. The second term would have been a step function that switches sharply from zero to one, if it were not due to preacceleration that smoothly flips the sign of the acceleration [Fig.~\ref{fig:Short}(a), red]. 
Integrating the above approximated acceleration for $\Delta T<0$, the rapidity [Fig.~\ref{fig:Short}(a), black] can be approximated by
\begin{equation}
    \label{eq:wn}
    w\simeq w_n+(-1)^n\epsilon (\Delta T+2-2e^{\Delta T}).
\end{equation}
Since the future force is in the opposite direction, the particle starts to decelerate before reaching $z=0$. Consequently, upon returning to the midplane, the particle is not able to regain the rapidity it had during the previous zero crossing. The rapidity lost can be found from Eq.~(\ref{eq:wn}), which gives
\begin{equation}
    \label{eq:adw}
    |w_{n-1}|-|w_n|\simeq 2\epsilon.
\end{equation}
While these intriguing behaviors occur for $\Delta T<0$, what happens for $\Delta T>0$ is much simpler. After crossing $z=0$, the particle sees a constant external force until $\Delta T\approx d_{n+1/2}$, when the particle again returns to the midplane. Since preacceleration is exponentially suppressed, the remote future has far less influence than the near future. When $d_{n+1/2}\gg 1$, to a very good approximation, we can ignore future zero crossings. Then, for $0<\Delta T\ll d_{n+1/2}$, 
\begin{eqnarray}
    \label{eq:dwp}
    \dot{w}&\simeq& -\epsilon (-1)^{n},\\
    \label{eq:wp}
    w&\simeq& w_n-\epsilon (-1)^n \Delta T.
\end{eqnarray}
 In other words, right after $T_n$, the particle is decelerated by an almost constant force and the motion is very close to hyperbolic. As its name suggests, preacceleration mostly affects how the particle approaches $z=0$, with little influence on how the particle leaves the midplane.

The influence of preacceleration can also be seen from the energy components [Fig.~\ref{fig:Short}(b)]. Denoting $T_{n-1/2}$ the time at which the particle reaches the maximum displacement where the potential energy is maximized [Fig.~\ref{fig:Short}(b), blue]. At that instant, the rapidity $w(T_{n-1/2})=0$, so both the kinetic and the Schott energy are zero. 
When $T_{n-1/2}<T<T_n$, the potential energy decreases while the kinetic energy increases [Fig.~\ref{fig:Short}(b), red]. Their sum, namely, the mechanical energy [Fig.~\ref{fig:Short}(b), magenta] is roughly constant until $T$ approaches $T_n$.
Near the bottom of the potential well, preacceleration acts as a friction force. Consequently, only a fraction of $\mathcal{E}_P$ is converted to $\mathcal{E}_K$, while the remaining energy is transferred to the Schott energy [Fig.~\ref{fig:Short}(b), orange]. 
Once the particle crosses the midplane, preacceleration plays a negligible role. The particle's kinetic energy is almost entirely converted to its potential energy, and the increasing radiation energy [Fig.~\ref{fig:Short}(b), purple] is almost entirely provided by the Schott energy. In other words, mechanical energy is consumed to charge the Schott energy only during zero crossings, while the Schott energy is continuously discharged to supply energy for radiation.

The amount of mechanical energy lost between $T_{n-1/2}$ and $T_{n+1/2}$ can be easily estimated. Denoting $\Delta\mathcal{E}=\mathcal{E}(T_{n+1/2})-\mathcal{E}(T_{n-1/2})$, then $\Delta\mathcal{E}_K=\Delta\mathcal{E}_S=0$ because both the kinetic and the Schott energy vanishes at maximum displacements.
For $\Delta T<0$, the energy change can be calculated using approximations Eqs.~(\ref{eq:dwn}) and (\ref{eq:wn}). Similarly, for $\Delta T>0$, the energy change can be calculated using approximations Eqs.~(\ref{eq:dwp}) and (\ref{eq:wp}). Also using these approximation, we have
\begin{eqnarray}
    \label{eq:qnn}
    T_n-T_{n-1/2}&\simeq& \frac{|w_n|}{\epsilon}+2,\\
    \label{eq:qnp}
    T_{n+1/2}-T_n &\simeq& \frac{|w_n|}{\epsilon},
\end{eqnarray}
which are needed as the upper and lower limits of the integrals.
Using either Eq.~(\ref{eq:Ep}) or Eq.~(\ref{eq:Er}), it is a straightforward calculation to show that
\begin{equation}
    \label{eq:dEhalf}
    \Delta\mathcal{E}_R=-\Delta\mathcal{E}_P\lesssim2\epsilon\sinh |w_n|,
\end{equation}
plus $O(\epsilon^2)$-order terms. 
We see that after each half period, the particle effectively looses its potential energy due to radiation. The energy change is larger for larger acceleration because the radiation power is proportional to $\epsilon^2$. Moreover, the energy change is larger when the particle has more time to radiate: the oscillation half period is $\sim 2\sinh |w_n|/\epsilon$ in the lab frame as I will show later [Eq.~(\ref{eq:Tbar})]. 
The above formula overestimates the increase of the radiation energy as well as the decrease of the potential energy.

\subsection{Long-time behavior}
Although changes within each oscillation period is small, the accumulated change can be substantial. 
To describe the long-time behavior, let us define an auxiliary envelope function $\bar{w}$, such that $\bar{w}(T_n)\simeq|w_n|$.
At intermediate time $T=T_{n-1/2}$, we can approximate the envelope function by $\bar{w}(T_{n-1/2})\simeq|w_{n-1}|-\epsilon\simeq|w_{n}|+\epsilon$, and approximate its time derivative by
\begin{equation}
    \label{eq:dwbar}
    \frac{d\bar{w}}{dT} =\frac{|w_n|-|w_{n-1}|}{d_{n-1/2}}
    \simeq-\frac{\epsilon^2}{\bar{w}+\epsilon}.
\end{equation}
Equivalently, the above differential equation for $\bar{w}$ can be derived by fitting the mechanical or radiation energy. Denoting $\bar{\mathcal{E}}_M$ the averaged mechanical energy, then the averaged radiation energy $\bar{\mathcal{E}}_R=\mathcal{E}-\bar{\mathcal{E}}_M$, because the averaged Schott energy $\bar{\mathcal{E}}_S=0$. Since the conserved total energy $\mathcal{E}$ is only shared between $\bar{\mathcal{E}}_M$ and $\bar{\mathcal{E}}_R$, we can impose the time derivatives at $T=T_n$ to be
\begin{equation}
    \label{eq:dEbar}
    \frac{d \bar{\mathcal{E}}_M}{dT}=-\frac{d \bar{\mathcal{E}}_R}{dT}=-\frac{\Delta \mathcal{E}_R}{T_{n+1/2}-T_{n-1/2}}\simeq \frac{\epsilon^2\sinh \bar{w}}{\bar{w}+\epsilon}.
\end{equation} 
To convert the above equation to an equation for $\bar{w}$, notice that at $T=T_n$, the potential energy $\mathcal{E}_P=0$, while the kinetic energy $\mathcal{E}_K=\cosh w_n-1$. Since $\bar{w}(T_n)\simeq|w_n|$, we can express the averaged mechanical energy as 
\begin{equation}
    \label{eq:w2E}
    \bar{\mathcal{E}}_M\simeq\cosh\bar{w}-1.
\end{equation}
Substituting this relation into Eq.~(\ref{eq:dEbar}), we obtain the same differential equation for $\bar{w}$ as given by Eq.~(\ref{eq:dwbar}). This differential equation is easy to solve:
\begin{equation}
    \label{eq:wbar}
    \bar{w}\simeq\sqrt{(w_0+\epsilon)^2-2\epsilon^2T}-\epsilon,
\end{equation}
which gives a good approximation of the long-time behavior when $w_0\gg\epsilon$ and $\epsilon\ll 1$.
We see the envelope decreases to zero in finite time $T_c\simeq(w_0+\epsilon)^2/2\epsilon^2$.
In other words, if we launch a charged particle in the electrostatic trap, the particle will undergo hyperbolic oscillations for only a finite time, before all its mechanical energy is drained by radiative energy loss. Of course, the above approximation is not valid near $T_c$, because when $T\rightarrow T_c$, the rapidity $w\rightarrow 0$; when $w$ is not much larger than $\epsilon$, the condition $d_{n-1/2}\gg1$ no longer hold. Therefore, the above lowest-order approximations deviate from the exact results when approaching the final time.

\begin{figure}[b!]
\centering
\includegraphics[width=1.0\linewidth]{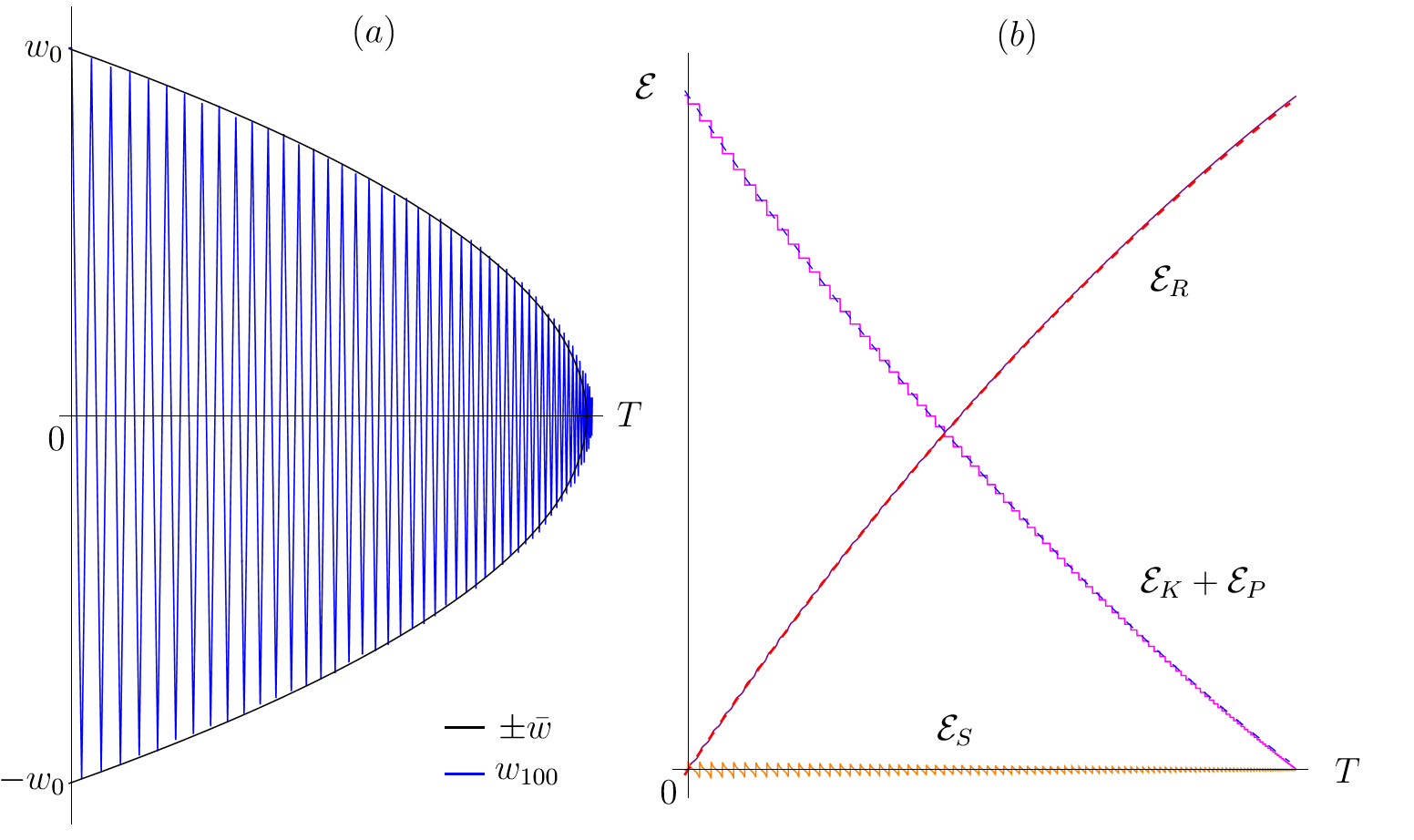}
\caption{Long-time behavior of the hyperbolic oscillator can be well approximated using the envelope function. For example, when $\epsilon=0.01$, $w_{N,N}=0.1$, and $N=100$, the exact solution to the final value problem (a, blue) is well inscribed by the envelope function (a, black). The stepwise decreasing mechanical energy (b, magenta) is centered around its envelope approximation (b, dashed blue), and the smoothly increasing radiation energy (b, purple) almost overlaps with is envelope approximation (b, dashed red). The Schott energy (b, orange) oscillates about zero, with no averaged contribution to the total energy.}
\label{fig:Long}
\end{figure}

An example of the long-time behavior of the trapped particle is shown in Fig.~\ref{fig:Long}. In this example, the normalized acceleration $\epsilon=0.01$, and the final value problem is solved for $N=100$ zero crossings with $w_{N,N}=0.1$. The exact solution of the rapidity [Fig.~\ref{fig:Long}(a), blue] shows hyperbolic oscillations with decreasing amplitude and increasing frequency. The black curves in Fig.~\ref{fig:Long}(a) shows the envelope $\pm\bar{w}$ given by Eq.~(\ref{eq:wbar}), which provides a very good approximation. 
Each time the particle crosses the midplane, the Schott energy [Fig.~\ref{fig:Long}(b), orange] is reset, which is then discharged to provide energy for radiation. The Schott energy oscillates and its averaged value is zero. 
On the other hand, the radiation energy keeps on increasing [Fig.~\ref{fig:Long}(b), purple].
Since the external acceleration has constant absolute value, the radiation power is almost constant. At early time when the particle has high rapidity, the radiation energy increases at a rate $\sim\epsilon^2\sinh \bar{w}/\bar{w}$, where the factor $\sinh \bar{w}/\bar{w}$ comes from the relativistic time dilation. At later time when $\epsilon\ll w\ll 1$, relativistic effects diminishes and the radiation energy increases linearly at a rate $\sim\epsilon^2$.
Since the total energy is conserved, the increase in radiation energy ultimately comes from the mechanical energy through the Schott energy. The mechanical energy, which is roughly constant away from zero crossings, decreases in a stepwise manner [Fig.~\ref{fig:Long}(b), magenta]. The average mechanical energy $\bar{\mathcal{E}}_M$, which is given by Eq.~(\ref{eq:w2E}), is shown by the dashed blue curve in Fig.~\ref{fig:Long}(b). 
When the acceleration $\epsilon$ is smaller, the approximation works even better.

\section{Experimental signatures: chirping of radiation \label{sec:chirp}}
Experimentally, it may be challenging to directly measure the trajectory and energy of a single particle. However, it is possible to measure the emitted radiation. 
Unlike a harmonic trap, the hyperbolic trap offers additional experimental signatures through radiation chirping.

The characteristic frequency of the electromagnetic wave is determined by the period of the hyperbolic oscillation. Since the oscillation period is slowly decreasing, one will observe positively chirped signals with increasing frequency.
To see how the frequency is chirped, we can use Eq.~(\ref{eq:dn}). The oscillation quarter period in the particle's frame is then
\begin{equation}
    \label{eq:T4}
    Q\simeq\frac{\bar{w}}{\epsilon}+1.
\end{equation}
Here, I relate the oscillation period to the envelope $\bar{w}$ because the period is well-defined only after averaging over many cycles.

Due to relativistic time dilation, the quarter period appears longer in the lab frame.  
The normalized lab time $\mathcal{T}=t/\tau_0$ between zero crossings can be determined using $d t/d\tau=\gamma=\cosh w$ as
\begin{equation}
    \label{eq:dbarn}
    \delta_{n-1/2}:=\mathcal{T}_n-\mathcal{T}_{n-1}=\int_{T_{n-1}}^{T_{n}}dT\,\cosh w.
\end{equation}
The exact value of $\delta_{n-1/2}$ can be approximated using Eq.~(\ref{eq:wn}) for $w$ when $T\in[T_{n-1/2},T_n]$ and using Eq.~(\ref{eq:wp}) for $w$ when $T\in[T_{n-1},T_{n-1/2}]$, where the time intervals can be approximated by Eqs.~(\ref{eq:qnn}) and (\ref{eq:qnp}). The oscillation half period in the lab frame is then
\begin{equation}
    \label{eq:dbar}
    \delta_{n-1/2}\simeq\frac{2\sinh |w_{n-1}|}{\epsilon}.
\end{equation}
From the half period, we can then determine $\mathcal{Q}$, the normalized lab-frame quarter period. Since the period is slowly drifting, we can define the quarter period at time $\mathcal{T}_n$ to be the average
\begin{equation}
    \label{eq:Tbarn}
    \mathcal{Q}_{n}:=\frac{\delta_{n-1/2}+\delta_{n+1/2}}{4}.
\end{equation}
To determine the smooth function $\mathcal{Q}$ that fits the above discrete points, we can again use the envelope function $\bar{w}$. Recall that at $T_n$ we have $\bar{w}=|w_n|\simeq|w_{n-1}|-2\epsilon$, so $\delta_{n-1/2}\simeq 2\sinh(\bar{w}+2\epsilon)/\epsilon$ and $\delta_{n+1/2}\simeq 2\sinh(\bar{w})/\epsilon$. In other words, we can express $\mathcal{Q}_{n}$ in terms of $\bar{w}(T_n)$. Enforcing this expression for all time, the lab-frame quarter period can then be approximated by
\begin{equation}
    \label{eq:Tbar}
   \mathcal{Q}\simeq\frac{\sinh(\bar{w}+\epsilon) \cosh\epsilon }{\epsilon}.
\end{equation}
It may be tempting to ignore $\epsilon$ in the numerator in comparison with $\bar{w}$. However, since $\sinh(\bar{w}+\epsilon)\simeq\exp(\bar{w})(1+\epsilon+\dots)$, the small $\epsilon$ term may still introduce order unity corrections.

To see how $\mathcal{Q}$ decreases in the lab time, we need to determine $\bar{w}$ as a function of the lab time $\mathcal{T}$. 
Following similar arguments that lead to Eq.~(\ref{eq:dwbar}), we can impose the derivative of $\bar{w}$ at $T_{n-1/2}$ to be
\begin{equation}
    \label{eq:dwbardTbar}
    \frac{d\bar{w}}{d\mathcal{T}} =\frac{|w_n|-|w_{n-1}|}{\delta_{n-1/2}}
    \simeq-\frac{\epsilon^2}{\sinh(\bar{w}+\epsilon)}.
\end{equation}
This equation can be easily integrated, whereby the rapidity envelope as a function of the lab time is given by
\begin{equation}
    \label{eq:wbarTbar}
    \cosh(\bar{w}+\epsilon) \simeq \cosh(w_0+\epsilon)-\epsilon^2\mathcal{T}.
\end{equation}
Notice that the envelope $\bar{w}$ decreases to zero in finite lab time. Moreover, $\mathcal{Q}$ decreases to zero when $\bar{w}=-\epsilon$, which is attained at the cutoff time 
\begin{equation}
    \label{eq:Tc}
    \mathcal{T}_c\simeq\frac{1}{\epsilon^2}\big[\cosh(w_0+\epsilon)-1\big].
\end{equation}
The above estimation of the final cutoff time has order unity error because the approximation is only valid when $\bar{w}\gg\epsilon$. The error is manifested in Eq.~(\ref{eq:Tbar}), where the oscillation period $\mathcal{Q}\simeq 1$ fails to decrease to zero even when the oscillation amplitude $\bar{w}=0$ has vanished.

The oscillation period decreases almost linearly for a long time. To see this, substituting Eq.~(\ref{eq:wbarTbar}) into Eq.~(\ref{eq:Tbar}), we can express the lab-frame quarter period in terms of the lab time as
\begin{eqnarray}
    \label{eq:linear}
    \mathcal{Q}&\simeq&\sqrt{\bigg(\mathcal{Q}_{0}-\frac{\epsilon \mathcal{T} \cosh\epsilon }{\tanh(w_0+\epsilon)}\bigg)^2-\bigg(\frac{\epsilon \mathcal{T} \cosh\epsilon}{\sinh(w_0+\epsilon)}\bigg)^2}\\
    \nonumber
    &\simeq& \mathcal{Q}_{0}-\frac{\epsilon\cosh\epsilon}{\tanh(w_0+\epsilon)}\mathcal{T}+O(\epsilon^2),
\end{eqnarray}
where $\mathcal{Q}_{0}=\sinh(w_0+\epsilon)\cosh\epsilon/\epsilon$ is the initial quarter period. 
The linear approximation holds when $\epsilon\mathcal{T}\ll \mathcal{Q}_{0}\sinh(w_0+\epsilon)/[1+\cosh(w_0+\epsilon)]$. In other words, the linear approximation holds for $\sim 1/\epsilon$ cycles of oscillations if $w_0\gtrsim 1$.
Another way to see the chirping rate is by directly computing
\begin{equation}
    \label{eq:dqbar}
    \frac{d\mathcal{Q}}{d\mathcal{T}}= \frac{d\mathcal{Q}}{d\bar{w}} \frac{d\bar{w}}{d\mathcal{T}}\simeq-\frac{\epsilon\cosh\epsilon}{\tanh(\bar{w}+\epsilon)}.
\end{equation}
We see the chirping rate $\sim \epsilon/\tanh\bar{w}$ is much smaller than the oscillation period $\sim\sinh \bar{w}/\epsilon$ as long as $\bar{w}\gg \epsilon$. Therefore, the radiated electromagnetic waves always have well-defined frequency, and the chirping rate is always slow within the applicability of classical electrodynamics.

\begin{figure}[b!]
\centering
\includegraphics[width=0.85\linewidth]{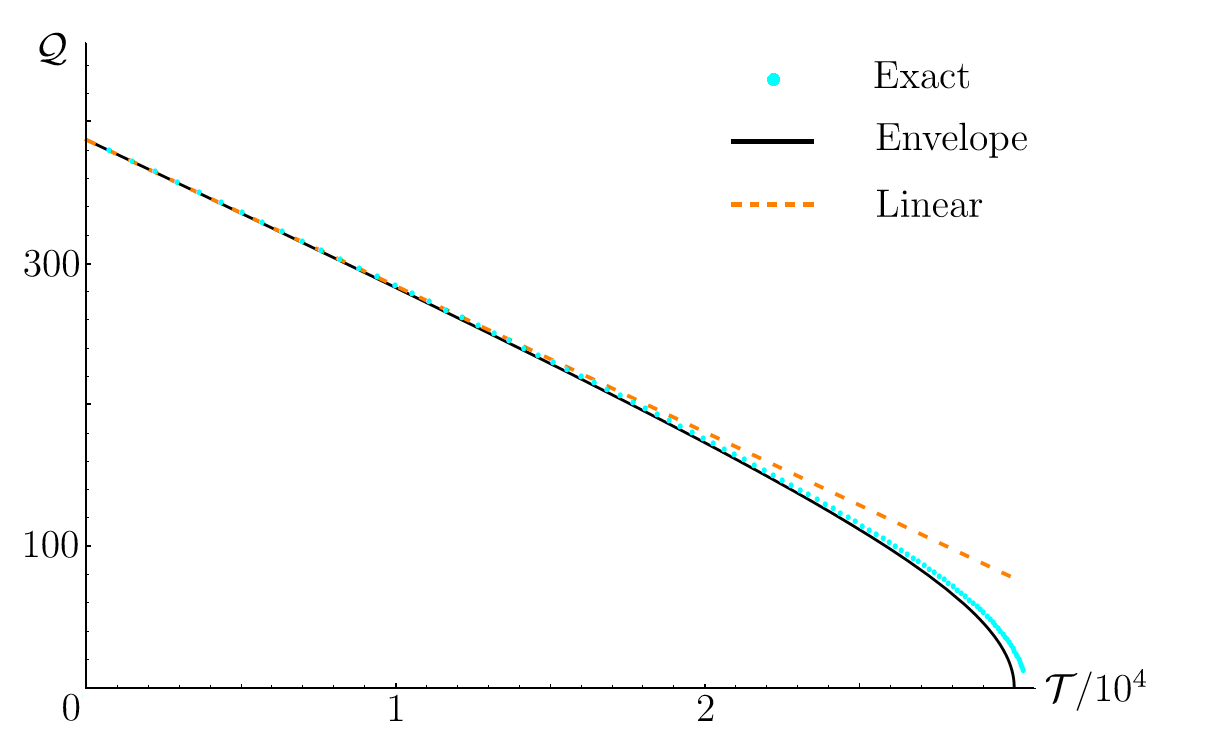}
\caption{In the lab frame, the oscillation quarter period $\mathcal{Q}$ decreases as the oscillation amplitude decays. When the lab time $\mathcal{T}\ll \sinh w_0/\epsilon^2$, the chirping is almost linear (dashed orange). At later time, the chirping is better estimated using an envelope approximation (black). Near the final cutoff time, the envelope approximation fails, and chirping is given by the exact solution (cyan dots). In this example, the exact solution is obtained by solving the final value problem with $\epsilon=0.01$, $w_{N,N}=0.1$, and $N=100$.}
\label{fig:Chirp}
\end{figure}

An example of the chirping behavior is shown in Fig.~\ref{fig:Chirp}. In this example, $\epsilon=0.01$, $w_{N,N}=0.1$, and $N=100$ are the same as in the previous example. The exact values of the lab-frame quarter period [Eq.~(\ref{eq:Tbarn})] as a function of the lab time can be determined by solving the final value problem.
Since $\epsilon\ll\bar{w}$, chirping can also be estimated using the envelope approximation [Eq.~(\ref{eq:Tbar})]. Moreover, in the same regime, the chirping is almost linear for a very long time [Eq.~(\ref{eq:linear})]. Suppose $\epsilon\ll\varepsilon\ll 1\lesssim w_0$ where $\varepsilon$ is some small number, then linear approximation holds for $\mathcal{T}\lesssim \mathcal{T}_{\text{max}}\simeq \varepsilon\sinh^2 (w_0+\epsilon)/\epsilon^2[1+\cosh(w_0+\epsilon)]$. Within this lab time $\mathcal{T}_{\text{max}}$, the particle oscillates for $\sim \varepsilon/\epsilon\gg1$ cycles, and the relative change of the period $\Delta \mathcal{Q}/\mathcal{Q}_{0}\sim\varepsilon$ can be significant. 
Recall that for classical electrodynamics to hold, we need $\epsilon\ll 10^{-2}$, so we can safely take $\varepsilon=0.1$. Then, if the particle initially has relativistic velocity $w_0\gtrsim 1$, we can observe as much as $\sim10\%$ chirping in the linear regime.

\section{Required experimental accuracy \label{sec:diff}}
To falsify classical models, experiments need to be accurate enough to resolve model differences. To estimate a typical difference, I consider the LL self force in addition to the LAD self force, both of which are widely advocated in the literature. It turns out that predictions based on the LL self force is equivalent to those given by the envelope approximation.

\subsection{Equivalence of Landau-Lifshitz and envelope approximations}
Using perturbative treatment, one first ignores the self force. 
Since the external force is constant except when the particle crosses the boundary, 
the time derivative of the lowest-order acceleration is 
\begin{equation}
    \label{eq:ddwLL}
    \ddot{w}^\text{LL}=-2\epsilon\sum_{n=-\infty}^{+\infty}(-1)^{n}\delta(T-T_n^\text{LL}),
\end{equation}
where the superscripts emphasize the LL approximation. The minus sign comes from the initial condition at $T_0$, when the particle enters the $z>0$ region where the acceleration is in the $-z$ direction.

The perturbative treatment then calculates the next-order acceleration by including the self force. This can be done by substituting Eq.~(\ref{eq:ddwLL}) into the LAD formula. When $T_{n-1}^\text{LL}<T<T_n^\text{LL}$, the LL self force is zero. The solution is simply hyperbolic motion:
\begin{equation}
    \label{eq:wLL}
    w^\text{LL}=w_{n-1}^+ +\epsilon(-1)^n(T-T_{n-1}^\text{LL}).
\end{equation}
Here, $w_{n-1}^+$ is the rapidity when $T$ approaches $T_{n-1}^\text{LL}$ from the positive side. What the LL self force does is to introduce discontinuities during zero crossings. Integrating across the $\delta$-function,
\begin{equation}
    \label{eq:wjump}
    w_{n-1}^+ =w_{n-1}^- +2\epsilon(-1)^n,
\end{equation}
where $w_{n-1}^-$ is the rapidity when $T$ approaches $T_{n-1}^\text{LL}$ from the negative side. Notice that $w_n=(-1)^{n}|w_n|$, so the $\delta$-function kick reduces the absolute value of the rapidity when the particle crosses the midplane.

\begin{figure}[t]
\centering
\includegraphics[width=1.0\linewidth]{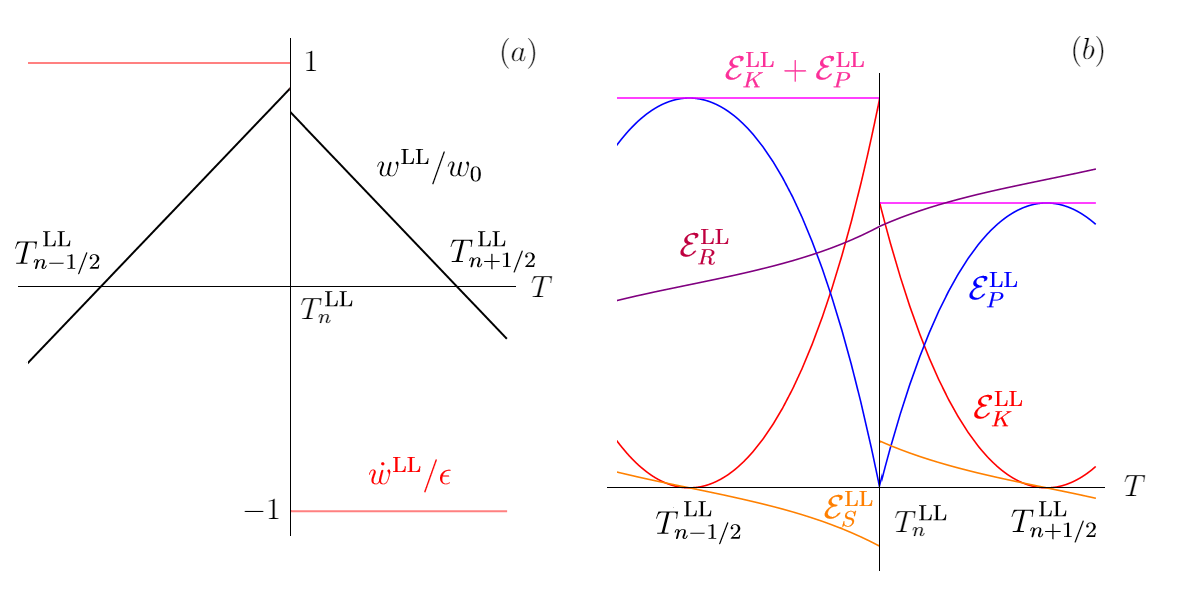}
\caption{Landau-Lifshitz approximation of the short-time dynamics of (a) the particle trajectory and (b) the energy components near $T_n^\text{LL}$. This figure is otherwise the same as Fig.~\ref{fig:Short}. The discontinuities at $T_n^\text{LL}$ are due to the $\delta$-function kick of the LL self force. }
\label{fig:ShortLL}
\end{figure}

As before, we do not know \textit{a priori} the zero crossing time, which needs to be determined from the particle's trajectory. Substituting Eq.~(\ref{eq:wLL}) into the constraint that the particle returns to zero [Eq.~(\ref{eq:constraint})], 
the time interval between zero corssings is 
\begin{equation}
    \label{eq:dnLL}
    d^\text{LL}_{n-1/2}=\frac{2|w_{n-1}^+|}{\epsilon}.
\end{equation}
We see the approximation for $d_{n-1/2}$ [Eq.~(\ref{eq:dn})] becomes exact under the LL self force. Substituting $T_n^\text{LL}=T_{n-1}^\text{LL}+d_{n-1/2}^\text{LL}$ into Eq.~(\ref{eq:wLL}) we can easily determine $w_n^-=-w_{n-1}^+$. 
In comparison with the previous zero crossing, the absolute value of the rapidity decreases by
\begin{equation}
    \label{eq:dwnp}
    |w_n^+|=|w_{n-1}^+|-2\epsilon.
\end{equation}
We see the short-time behavior given by Eq.~(\ref{eq:adw}) becomes exact under the LL self force. Notice that each time the particle returns to the midplane, the rapidity $|w|$ decreases by a fixed amount. However, the time it takes for the particle to return is gradually shrinking.
The short-time dynamics is shown in Fig.~\ref{fig:ShortLL}(a), where I have used exactly the same parameters as in Fig.~\ref{fig:Short}. Unlike the LAD self force that introduces a smooth decrease through preacceleration, the LL self force introduces jumps when the particle crosses the midplane.

To see how the LL self force affects the energy of the charged particle, let us calculate energy components on the time interval $T\in(T_{n-1}^\text{LL}, T_n^\text{LL})$. The expressions are particularly simple in terms of $T^\text{LL}_{n-1/2}$, the time when the particle reaches the maximum displacement where the rapidity becomes zero. It is easy to see that $T^\text{LL}_{n-1/2}=(T^\text{LL}_{n-1}+T^\text{LL}_{n})/2$ is exactly at the center of the time interval, on which $w^{\text{LL}}(T)=\epsilon(-1)^n(T-T^\text{LL}_{n-1/2})$. 
First, the kinetic energy is still given by Eq.~(\ref{eq:EK}), which can now be written as
\begin{equation}
    \label{eq:EKLL}
    \mathcal{E}_K^{\text{LL}}=\cosh\epsilon(T-T^\text{LL}_{n-1/2})-1.
\end{equation}
Since the rapidity is discontinuous, the kinetic energy is also discontinuous across time boundaries.
Second, the potential energy is given by Eq.~(\ref{eq:Ep}). The integration can now be carried out exactly: 
\begin{equation}
     \label{eq:EpLL}
    \mathcal{E}_P^{\text{LL}}=\cosh|w^+_{n-1}|-\cosh\epsilon(T-T^\text{LL}_{n-1/2}).
\end{equation}
The potential energy is clearly continuous and is in fact zero at the midplane. 
Third, the radiation energy, which is given by the integral Eq.~(\ref{eq:Er}), is also continuous. Denoting $\mathcal{E}_{R,n-1}^\text{LL}$ the value of the radiation energy at the previous time boundary, then
\begin{equation}
    \label{eq:ErLL}
    \mathcal{E}_R^{\text{LL}}=\mathcal{E}_{R,n-1}^\text{LL}+\epsilon\sinh|w^+_{n-1}|+\epsilon\sinh\epsilon(T-T^\text{LL}_{n-1/2}).
\end{equation}
We can compute $\mathcal{E}_{R,n}^{\text{LL}}$ from $|w^+_{n}|=w^+_{0}-2\epsilon n$ by carrying out the sum 
$\mathcal{E}_{R,n}^{\text{LL}}=2\epsilon \sum_{k=0}^{n-1}\sinh(w^+_{0}-2\epsilon k)=\epsilon[\cosh(w^+_{0}+\epsilon) -\cosh(w^+_{n}+\epsilon)]/\sinh\epsilon$,
where I have set the reference value $\mathcal{E}_{R,0}^{\text{LL}}$ to zero.
Finally, the Schott energy [Eq.~(\ref{eq:Es})] is now given by 
\begin{equation}
    \label{eq:EsLL}
    \mathcal{E}_S^{\text{LL}}=-\epsilon\sinh\epsilon(T-T^\text{LL}_{n-1/2}).
\end{equation}
Similar to the kinetic energy, the Schott energy is discontinuous across time boundaries. The short-time dynamics of all energy components are shown in Fig.~\ref{fig:ShortLL}(b), where I have used exactly the same parameters as in Fig.~\ref{fig:Short}. While the kinetic energy suddenly drops, the Schott energy suddenly increases due to the $\delta$-function kick by the LL self force.

Although the LL approximation results in a much simpler solution forward in time, it is important to note that the total energy is no longer conserved. This can be anticipated from the time derivative of the total energy Eq.~(\ref{eq:dE}), which is zero within each time interval but nonzero at the time boundaries.
To see how much energy error is introduced by the LL approximation, let us compare the total energy in two consecutive time intervals. Summing Eqs.~(\ref{eq:EKLL})-(\ref{eq:EsLL}), the total energy when $T_{n-1}^\text{LL}<T< T_n^\text{LL}$ is 
\begin{equation}
    \label{eq:EtLL}
    \mathcal{E}^\text{LL}_{n-1/2}=\cosh |w_{n-1}^+| -1+\mathcal{E}_{R,n-1}^\text{LL}+\epsilon\sinh|w_{n-1}^+|.
\end{equation}
Similarly, we can compute the total energy $\mathcal{E}^\text{LL}_{n+1/2}$ in the next time interval $T_{n}^\text{LL}<T< T_{n+1}^\text{LL}$. The energy difference $\Delta\mathcal{E}^\text{LL}_{n}=\mathcal{E}^\text{LL}_{n+1/2}-\mathcal{E}^\text{LL}_{n-1/2}$ can be computed using Eq.~(\ref{eq:dwnp}) to relate $w_{n-1}^+$ to $w_{n}^+$:
\begin{eqnarray}
    \label{eq:dEtLL}
    \nonumber
    \Delta\mathcal{E}^\text{LL}_{n}&=& \cosh|w_n^+|(1+\epsilon\sinh 2\epsilon-\cosh2\epsilon)\\
    &+& \sinh|w_n^+|\,(\epsilon+\epsilon\cosh 2\epsilon-\sinh2\epsilon).
\end{eqnarray}
The first line is proportional to $2\epsilon^4/3+O(\epsilon^6)$ and the second line is proportional to $2\epsilon^3/3+O(\epsilon^5)$.
The energy error at each time step is small and slowly decreasing. However, notice that $\Delta\mathcal{E}^\text{LL}_{n}>0$ is always positive for all values of $\epsilon$ and $w_n^+$. In other words, using the LL approximation, the total energy monotonically increases due to errors.
As an estimate of an upper bound of the accumulated error, $\Delta\mathcal{E}^\text{LL}=\sum_{n=1}^{w_0/2\epsilon}\Delta\mathcal{E}^\text{LL}_{n}\lesssim\epsilon^2w_0\sinh w_0/3$. When $\epsilon\ll w_0\lesssim 1$, the accumulated energy error is of $\epsilon^2$ order.

Having discussed properties of the LL solution, let me now show that the LL solution is exactly described by the envelope equation. To see this, we can identify 
$\bar{w}(T_{n-1/2}^\text{LL})=(|w_{n-1}^+| +|w_n^+|)/2$. Using Eq.~(\ref{eq:dnLL}) for the time interval $d_{n-1/2}^\text{LL}$ and Eq.~(\ref{eq:dwnp}) for the change of $|w|$, the time derivative of $\bar{w}$ is then exactly given by Eq.~(\ref{eq:dwbar}).
To see that the solution to the envelope equation [Eq.~(\ref{eq:wbar})] indeed passes through $|w_n^+|$, we can prove by mathematical induction. First, when $T=T_1^{\text{LL}}=d_{1/2}^\text{LL}$, Eq.~(\ref{eq:wbar}) gives $(\bar{w}_1+\epsilon)^2=(w_0+\epsilon)^2-4\epsilon|w_0^+|$. If we take the initial condition $w_0=|w_0^+|$, then $\bar{w}_1=|w_0^+|-2\epsilon=|w_1^+|$.
Next, suppose $\bar{w}_n=|w_n^+|$, namely, $(|w^+_n|+\epsilon)^2=(w_0+\epsilon)^2-2\epsilon^2T_{n}^\text{LL}$. 
Then, at the next time step $T_{n+1}^{\text{LL}}=T_n^{\text{LL}}+d_{n+1/2}^\text{LL}$, Eq.~(\ref{eq:wbar}) gives
\begin{eqnarray}
    \bar{w}_{n+1} =|w^+_{n+1}|,
\end{eqnarray}
We see the induction hypothesis also holds at the next step. Therefore, the continuous envelope $\bar{w}$ exactly passes through the LL solution $|w_n^+|$ at $T_{n}^\text{LL}$ for all $n$, and 
the two approximations are thereof equivalent.

Consequently, under the LL approximation, the chirping behaviors described by the envelope approximations also become exact. 
In particular, the proper quarter period $Q_{n}=(d_{n-1/2}+d_{n+1/2})/4$ is exactly given by Eq.~(\ref{eq:T4}). 
In terms of the lab time, Eq.~(\ref{eq:dbar}) becomes exact for $\delta_{n-1/2}$ if we identify $w_{n-1}$ with $w_{n-1}^+$.
Then, the lab frame quarter period $\mathcal{Q}$ is exactly given by Eq.~(\ref{eq:Tbar}) as a function of $\bar{w}$.
When expressing $\mathcal{Q}$ as a function of the lab time $\mathcal{T}$, 
Eq.~(\ref{eq:dwbardTbar}) becomes exact for $\bar{w}(\mathcal{T})$, whose solution Eq.~(\ref{eq:wbarTbar}) also becomes exact.
Substituting $\bar{w}(\mathcal{T})$ into $\mathcal{Q}(\mathcal{T})$, the lab frame qurter period can then be expressed in terms of the lab time, for which Eq.~(\ref{eq:linear}) becomes exact.
The LL approximation overestimates radiation energy and underestimate mechanical energy. The net consequence is that the rapidity and the quarter period are underestimated, while the total energy increases secularly.

\subsection{Discrepancies between envelope and exact solutions}
Neither the LL solution nor the preacceleration solution gives a satisfactory prediction of what will happen to a charged hyperbolic oscillator. 
The LL solution cannot be the true, because it does not even conserve energy. The preacceleration solution, albeit conserving energy and momentum, is usually regarded as unphysical because it violates classical causality. The real behavior of the hyperbolic oscillator is likely influenced by quantum effects, especially during midplane crossings that occur on Compton time scale. Although it remains to be calculated quantum mechanically and measured experimentally the long-time behavior of a trapped particle, it worth taking a closer look at the discrepancies between the two classical solutions.

\begin{figure}[b!]
\centering
\includegraphics[width=1.0\linewidth]{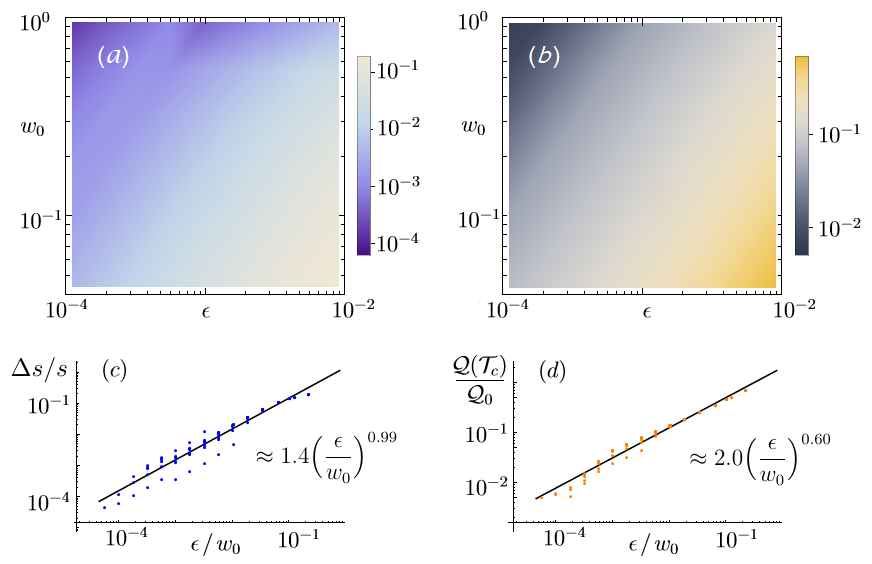}
\caption{Discrepancies between the LAD preacceleration solution and the Landau-Lifshitz solution give estimates of experimental accuracies needed to exclude radiation reaction models. The relative differences of (a) the linear chirping rate $\Delta s/s$  and (b) the qurter period $\mathcal{Q}(\mathcal{T}_c)/\mathcal{Q}_{0}$ at LL cutoff are both significant for large acceleration $\epsilon$ and small initial rapidity $w_0$. The discrepancies roughly scale as $\sim\sigma(\epsilon/w_0)^r$, where the error coefficient $\sigma$ and scaling exponent $r$ may be determined using linear regressions as shown in (c) and (d). The data points used in regressions are obtained by numerically solving the final value problem at $\sim 100$ uniformly distributed sampling points in the $\epsilon$-$w_0$ parameter space. }
\label{fig:Para}
\end{figure}

The discrepancies between the preacceleration and the LL solutions give estimates of what experimental accuracy will be needed in order to falsify classical radiation reaction models. 
At earlier time, the discrepancy can be quantified by measuring the linear chirping rate of the radiated electromagnetic waves. Linear chirping remains a very good approximation up to $\mathcal{T}_\text{max}$. 
We can thereof fit the exact $w_n$ linearly up to $n_\text{max}\simeq\varepsilon\sinh (w_0+\epsilon)/\epsilon[1+\cosh(w_0+\epsilon)]$, and compare the exact slope with the slope given by the envelope approximation [Eq.~(\ref{eq:linear})]. The exact slope is more negative than the envelope slope because the envelope approximation is a convex interpolation of the exact data. Denoting $s$ the exact slope and $s^\text{LL}$ the slope given by LL approximation, the relative difference $\Delta s/s$ at $\mathcal{T}_0$ is shown in Fig.~\ref{fig:Para}(a). 
As expected, the discrepancy is larger for larger $\epsilon$ and smaller $w_0$, whereby the oscillation period is closer to the Compton time.

At later time, the linear approximation breaks down and the discrepancy between the two solutions increases monotonically. The discrepancy becomes most pronounced at $\mathcal{T}_c$ [Eq.~(\ref{eq:Tc})], when $\mathcal{Q}^\text{LL}$ given by the envelope solution is reduced to zero while the exact quarter period is still finite. In other words, the LL solution predicts that the particle will lose all its mechanical energy and come to rest with no further radiation at $\mathcal{T}_c$, while the preacceleration solution predicts that the particle is still in motion and continues to radiate beyond $\mathcal{T}_c$. The ratio $\mathcal{Q}(\mathcal{T}_c)/\mathcal{Q}_{0}$ is shown in Fig.~\ref{fig:Para}(b), where $\mathcal{Q}(\mathcal{T}_c)$ is the quarter period of the exact solution at the LL cutoff and $\mathcal{Q}_{0}$ is the initial quarter period.
Since the exact data is discrete, $\mathcal{Q}(\mathcal{T}_c)=(q_L \delta_R+q_R \delta_L)/(\delta_L+\delta_R)$ is estimated using linear interpolation, where $q_L=\mathcal{Q}_{n_c-1}$ and $q_R=\mathcal{Q}_{n_c}$ are the nearest exact quarter periods on the left and right of $\mathcal{T}_c$. The linear average is weighted by the time separations $\delta_L=\mathcal{T}_c-\mathcal{T}_{n_c-1}$ and $\delta_R=\mathcal{T}_{n_c}-\mathcal{T}_{c}$.
As shown in Fig.~\ref{fig:Para}(b), the discrepancy is again larger where quantum corrections are expected to become important.

When $w_0$ is not much larger than $\epsilon$, it is numerically affordable to compute the preacceleration solution. 
This exact solution is obtained by solving the final value problem with $w_{N,N}=\epsilon$ for $N\simeq w_i/2\epsilon$ steps, where $w_i$ is the targeted initial rapidity. Since the envelope approximation overestimates $w$ for the final value problem, the preacceleration solution $w_{N,0}$ is smaller than $w_i$. Given $w_{N,0}$ as the initial value, the LL solution, which can be easily obtained using the envelope formulas, then terminates after $n_c<N$ zero crossings. The discrepancies in the $\epsilon$-$w_0$ parameter space, as shown in Fig.~\ref{fig:Para}, are plotted using the actual value of the initial rapidity.

When $w_0$ is much larger than $\epsilon$, it is not always numerically feasible to solve the final value problem. In order to fit the slope and search the cutoff points, we need to store information at all $N\sim w_0/\epsilon$ time steps in the computer memory, which becomes very challenging for realistic $\epsilon$ values even on modern computers.
In this case, the discrepancies may be extrapolated using the empirical scaling laws shown in Fig.~\ref{fig:Para}(c) and (d). 
Although the discrepancies are not simple functions of $\epsilon/w_0$, their bulk parts may be grossly estimated as $\sim\sigma(\epsilon/w_0)^r$. By calculating the discrepancies at $\sim 100$ sampling points in the $\epsilon$-$w_0$ parameter space, the error coefficient $\sigma$ and the scaling exponent $r$ are estimated.
For initial linear chirping, the fitted parameters are $\sigma_c\approx 1.411$ and $r_c\approx 0.987$; for final quarter period, the fitted parameters are $\sigma_q\approx 1.991$ and $r_q\approx 0.602$.
To exclude classical radiation reaction models, the experimental uncertainty must be smaller than the expected discrepancies between the LL and the preacceleration solutions.

\section{Significance of quantization \label{sec:quantization}}
So far, I have assumed that classical electrodynamics is applicable to the hyperbolic oscillator, at least when the oscillation period is much longer than the Compton time. But, is classical electrodynamics really valid?

A prerequisite for applying classical electrodynamics is that the emitted radiations are continuous electromagnetic waves. In other words, if the radiations are in the form of discrete photons, classical electrodynamics may no longer hold. 
Since the hyperbolic oscillator radiates with increasing frequency and decreasing amplitude, the rate at which photons are emitted diminishes. 
To estimate the number of radiated photons $\mathcal{N}$ within each oscillation period, we can write $\Delta E_R=\hbar\omega\mathcal{N}$. The radiation energy per period is $\Delta E_R\simeq 4mc^2\epsilon\sinh\bar{w}$, where I have used Eq.~(\ref{eq:dEhalf}) for $\Delta\mathcal{E}_R$. The photon frequency is $\omega=\pi/2\tau_0\mathcal{Q}$, where the normalized quarter period $\mathcal{Q}\simeq\sinh\bar{w}/\epsilon$ is given by Eq.~(\ref{eq:Tbar}). The radiated photons per period is then
\begin{equation}
    \label{eq:Nphotons}
    \mathcal{N}\simeq\frac{16\alpha}{3\pi}\sinh^2\bar{w},
\end{equation}
where $\alpha$ is again the fine structure constant. We see the number of photons is less than one, unless the rapidity $\bar{w}\gg 1$. In other words, unless the oscillator has relativistic energy, it cannot radiate electromagnetic waves continuously! 
This is also true for a simple harmonic oscillator, for which $\mathcal{N}\simeq 4\pi\alpha\beta^2/3\ll 1$ in the nonrelativistic limit.
In experiment, a nonrelativistic charged particle will oscillate without radiation for many periods and then stochastically emit a photon only once in a while. Nevertheless, the photon rate may be quite large because of the large oscillation frequency, and the time-averaged quantum emission may still be approximated by the classical description.

Although radiation is almost always quantized, quantization of the charged particle may not always be important.
In what follows, let us estimate energy levels of the quantum charged particle in the absence of radiation. Since spin is not essential here, it is sufficient to solve the Klein-Gordon equation for the relativistic particle. In the Coulomb gauge, using the natural units $\hbar=c=1$, the wave function $\psi$ satisfies
\begin{equation}
    [(\partial_t-ie\phi)^2-\nabla^2+m^2]\psi=0,
\end{equation}
where the background potential $\phi=E|z|$. Since the problem is separable, we can write the wave function as $\psi=\varphi(z)\xi(x,y)\exp(iHt)$, where the frequency $H>m$ for particle states. Then, $\varphi$ satisfies the equation
\begin{equation}
    [\partial_z^2+V(z)]\varphi=0,
\end{equation}
where the effective potential $V=(H-e\phi)^2-m^2$ has four turning points at $|z^\pm_T|=(H\pm m)/eE$. Beyond $z^+_T$, the potential energy is sufficient to create electron-positron pairs. This is not the kind of quantum effects one will typically encounter in a feasible electrostatic trap.

Let us instead focus on nonrelativistic quantum effects when the charged particle crosses the midplane. Such quantum effects will be larger when the particle spends a larger fraction of its oscillation period near the midplane. In this case, the energy $H=m+K$ is only slightly above the rest energy. Since the turning points $z_T^+\gg z_T^-$ are well separated, the Bohr-Sommerfeld quantization condition can be well approximated by
\begin{eqnarray}
    \label{eq:BSquant}
    (n+\frac{1}{2})\pi\simeq\int_{-z^-_T}^{+z^-_T}V^{1/2}dz
    \simeq\frac{2K^2}{eE} \frac{2}{3}\sqrt{\frac{2m}{K}},
\end{eqnarray}
where I have taken the nonrelativistic limit $K\ll 2m$. The quantized kinetic energy is then the same as given by the Schr\"odinger's equation:
\begin{equation}
    \label{eq:Kn}
    K_n\simeq\frac{mc^2}{2}\Big[\frac{3\pi E}{2E_c}\Big(n+\frac{1}{2}\Big)\Big]^{2/3},
\end{equation}
where $n=0,1,\dots$ takes integer values. 
Here, $E_c=m^2c^3/e\hbar$ is again the Schwinger critical field for QED, and I have restored factors of $\hbar$ and $c$ for clarity. We see the above nonrelativistic approximation holds only if $E\ll E_c$ as expected.

Although it may seem that the energy levels approach continuum for large $n$, it is important to note that in the nonrelativistic limit, $n$ is bounded from above by the initial energy of the particle. 
When $1\ll n\ll E_c/E$, the energy gap between adjacent levels is well approximated by 
\begin{equation}
    \Delta_{n-1/2}:=K_n-K_{n-1}\simeq mc^2 \Big(\frac{\pi E}{2E_c}\Big)^{2/3}\Big(\frac{1}{3n}\Big)^{1/3}.
\end{equation}
Now, suppose we release the particle from rest at $z_0$, the initial mechanical energy of the particle is $K_i=eEz_0$. Substituting $K_i$ into Eq.~(\ref{eq:BSquant}), we can estimate the highest occupation level $n_i$. 
The initial quantization is then
\begin{equation}
    \Delta_i\simeq\frac{\pi\hbar}{2}\Big(\frac{eE}{2mz_0}\Big)^{1/2}.
\end{equation}
To see how does this initial quantization compare with the initial photon energy,
we can use Eq.~(\ref{eq:Tbar}) to estimate the normalized quarter period $\mathcal{Q}_0\simeq\sinh w_0/\epsilon$. Since the initial rapidity is such that $\cosh w_0=1+K_i/mc^2$, we have $\sinh w_0=\sqrt{\cosh^2 w_0-1}\simeq \sqrt{2 K_i/mc^2}$ in the nonrelativistic limit. The energy of the initial photon is then
\begin{equation}
    \hbar\omega_0=\frac{2\pi\hbar}{4\tau_0\mathcal{Q}_0}\simeq\frac{\pi\hbar}{2}\Big(\frac{eE}{2mz_0}\Big)^{1/2}.
\end{equation}
We see the photon energy is exactly the same as the initial quantization. That the classical calculation agrees with the quantum calculation may not be too surprising. In the case of a simple harmonic oscillator, the energy quantization $\hbar\omega$ also equals to the energy of the radiated photons.

The classical point-charge picture is resembled by the coherent state rather than eigenstates of the quantum system. The coherent state subsumes a collection of photons that the particle can emit. The classical picture is valid when this collection contains many photons. This will be the case when the mechanical energy of the particle is much larger than the energy quanta. The classicality condition that $K\gg \Delta$ can be written as
\begin{equation}
    \frac{2}{\pi}\Big(\frac{z}{\lambdabar}\Big)^{3/2}\Big(\frac{2E}{E_c}\Big)^{1/2}\gg 1,
\end{equation}
where $\lambdabar=\hbar/mc$ is the Compton wavelength. Since we have assumed $E\ll E_c$, the motion is classical only when the maximum displacement $z$ is large enough.
The above condition can be translated to a condition for the rapidity using $eEz/mc^2=K=\cosh\bar{w}-1\simeq \bar{w}^2/2$. The classicality condition for the rapidity is then
\begin{equation}
    \label{eq:wquantum}
    \bar{w}\gg \Big(\frac{\pi E}{E_c}\Big)^{1/3}=\Big(\frac{3\pi\epsilon}{2\alpha}\Big)^{1/3}.
\end{equation}
When the above condition is satisfied, the oscillation period is necessarily much longer than the Compton time, which agrees with our earlier intuition.
However, for the hyperbolic oscillator, the oscillation period decays with the oscillation amplitude, and quantum effects will always become important.

\section{Discussion and Summary \label{sec:conclusion}}
In an electrostatic trap of macroscopic size, one would expect that the behavior of a charged particle be well described by classical electrodynamics when the restoring electric field is far below the Schwinger field. This is a reasonable expectation because when the size of the trap is much larger than the characteristic size of the wave function, the charged particle is point-like; and when the restoring electric field is much smaller than the Schwinger field, quantum electrodynamics introduces little correction. 
However, as I have shown in this paper, quantization is more easily important, because the radiation power is so low compared to the oscillation frequency that the radiation must be emitted in the form of discrete photons, unless the charged particle has relativistic energy.

If we nevertheless apply classical electrodynamics as some sort of averaged description of the quantum system, what we then predict is that the charged particle will radiate electromagnetic waves continuously. The far-field electromagnetic waves will propagate away from the particle and bring away energy and momentum. 
The charged particle should then dissipate its mechanical energy in order for the total energy to be conserved. The global energy conservation requires a local mechanism. 
The local mechanism is provided by the self force, namely, the force a particle feels due to its own electromagnetic fields.
In classical electrodynamics, the self force turns out to act on $\tau_0$ time scale, which is much shorter than the Compton time.

Now we face a dilemma. On one hand, if we ignore radiation, we will predict that the charged particle oscillates with constant amplitude. In this lowest-order approximation, classical electrodynamics is valid, but its prediction is clearly wrong. On the other hand, if we include radiation, we will predict that the charged particle oscillates with diminishing amplitude. Although this next-to-leading order prediction is qualitatively correct, it involves applying classical electrodynamics on scales for which it is not valid. So, can we trust classical electrodynamics at all? What will really happen to a charged particle in an electrostatic trap?

While the classical dilemma may be hiding when the confining potential is a smooth function, it becomes manifested when the acceleration has sudden jumps. In the case of a smooth confining potential, one may perform Landau-Lifshitz reduction of order to remove the time derivative of acceleration in the LAD self force. This perturbative technique becomes more valid when radiation reaction is less important. In other words, when the experimental setup is such that radiation reaction is important for the dynamics of the charged particle, the LL approximation becomes less accurate. 
Moreover, the LL treatment
becomes singular when the confining potential has discontinuities. 
In this case, the presumably-small self force in fact gives $\delta$-function kicks each time the charged particle crosses the discontinuity.
While this may work fine mathematically, the presence of the delta function implies that the self force acts on infinitesimal time scales, for which classical electrodynamics is doomed to fail. In the case of hyperbolic oscillator, the failure is obvious from the secular increase of the total energy error.

The failure of classical electrodynamics is less catastrophic if we directly solve the Newton's equation including the LAD self force without performing the reduction of order. In this paper, I have shown how this can be achieved in great details. The physical solution, free from the pathological runaway behavior, can only be obtained when we trade runaway for preacceleration. Preacceleration, which is usually regarded as another pathology of the LAD self force, turns out to result in a reasonable solution that is similar to the LL solution. 
In fact, preacceleration makes physical sense if the charged particle is of finite size. This is because a finite-size particle can already feel the force on the other side of the midplane, even when its center of mass has not yet crossed the boundary.
In reality, charged particles are quantum, whose wave function has finite extent. 
Moreover, a quantum wave packet is not a rigid body. The wave packet can undergoes compression
when the particle crosses the midplane, where mechanical energy is consumed to charge the Schott energy. Subsequently, the wave packet can undergo expansion during the hyperbolic motion, where the stored Schott energy is discharged through radiation.

Of course, a heuristic quantum interpretation cannot justify the preacceleration solution, and
the question is whether the classical prediction is quantitatively correct? In terms of experimental observables, classical electrodynamics predicts that the radiated electromagnetic waves are chirped in a specific way. The questions is then whether the predicted chirping matches what really happens in an experiment?

As an example, let us consider an plausible experiment in which the restoring electric field is \mbox{$E\approx$ 1 MeV/m}. This electric field is large but not yet large enough to breakdown dry air and is experimentally feasible. In this electric field, the normalized acceleration $\epsilon=2\alpha E/3E_c \approx 3.7\times 10^{-15}$. Suppose the electrostatic trap is built with half size \mbox{$z_0\approx 1$ cm}, then the normalized potential energy is $\mathcal{E}_P=\epsilon z/\tau_0c\approx 0.02$ at the maximum displacement. If we launch the particle at rest from $z_0$, then the initial rapidity $w_0=\cosh^{-1}(1+\mathcal{E}_P)\approx0.2$. The normalized initial quarter period is $\mathcal{Q}_0\simeq\sinh w_0/\epsilon\approx 5.4\times 10^{13}\gg 1$, which corresponds to radiations with frequency $f_0=1/4\mathcal{Q}_0\tau_0\approx$ \mbox{737.8 MHz} in the radio range. The initial photon energy is about \mbox{3 $\mu$eV}, which is much smaller than the mechanical energy of the electron. Therefore, classical description of the electron is expected to be valid at the early stage of the experiment.

Once we launch the electron in the trap, the oscillating charge will dominantly emit radiation if the ambient temperature is colder than \mbox{$\sim 10$ mK}, which is attainable inside dilution refrigerators.
The initial linear chirping rate is $d\mathcal{Q}/d\mathcal{T}\simeq-\epsilon/\tanh w_0\approx-1.9\times 10^{-14}$, which remains a good approximation up to $t_\text{max}\sim0.1\times\tau_0\sinh^2w_0/\epsilon^2\sim$ \mbox{$ 907$ s}. At the end of the linear stage, the LL approximation predicts that the particle will radiate at \mbox{$f_l\approx$ 778.1 MHz}, which differs noticeably from $f_0$ by about $5\%$. However, the LL and the LAD predictions differ by only \mbox{$\sim 10^{-5}$ Hz}, which is too minuscule to be observable.

To see larger discrepancies, one can wait for longer time. The longest time one can wait is $t_c=\tau_0\mathcal{T}_c\approx$2.5 hours. At this cutoff time, LL approximation predicts that the oscillation frequency becomes infinite. A less ridiculous prediction, made by extrapolating the preacceleration solution, is still as high as \mbox{$\sim 10^{16}$ Hz}, which corresponds to photon energy of \mbox{$\sim 100$ eV}. 
Radiation with such high frequency is apparently outside the applicability of classical electrodynamics.  In fact, when approaching the cutoff, the mechanical energy of the electron will approach the ground state energy \mbox{$K_0\approx 3.75$ meV}. It is clear that the electron will not have sufficient energy to radiate the high-energy photons expected from classical electrodynamics.

Now the question is: before the classical description of the electron fails, will the LL and the preacceleration solutions already accumulate sufficient discrepancy to be discernible?
In the aforementioned experimental conditions, we need $\bar{w}\gg 10^{-4}$ for point-particle picture to hold. The highest radiation frequency is then \mbox{$\sim 10^{12}$ Hz}, which corresponds to photon energy \mbox{$\sim 1$ meV}.
Using the envelope approximation, the time $t_q$ when the oscillator enters the quantum regime turns out to be ahead of $t_c$ by only \mbox{$\sim 1$ ms}.
At $t_q$, the quarter period given by the LL solution is $\mathcal{Q}^\text{LL}_q\sim10^{10}$. The quarter period given by the preacceleration solution will be longer, but the discrepancy $\Delta\mathcal{Q}<\mathcal{Q}(\mathcal{T}_c)\sim10^5$.
We see the discrepancy is at most \mbox{$\sim$ 10 ppm}. In other words, when measuring terahertz radiations at the final stage of the experiment, the detector needs to have resolution finer than $\Delta f\sim10$ MHz and timing accuracy better than $\Delta t\sim 1$ ms, in order to discern the two classical radiation reaction models, if they could indeed provide some averaged descriptions of the \textit{de facto} quantum system.

In summary, it remains to be tested experimentally whether classical electrodynamics is applicable when radiation reaction is important. The trap-type experiments considered here is complementary to the collider-type experiments currently under investigation. In this paper, I have discussed hyperbolic trap in great details, where an electron is confined between two regions of constant but opposite accelerations. 
The classical electron travels with almost constant mechanical energy within each region and the radiation energy is provided almost entirely from the Schott energy, which is recharged by the mechanical energy only when the electron crosses the midplane.
The oscillating electron will radiate electromagnetic waves with diminishing amplitude and increasing frequency. The frequency chirping provides a distinct signature, using which classical radiation-reaction models may be tested experimentally. 
%



\section*{Acknowledgement}
This work was performed under the auspices of the U.S. Department of Energy by Lawrence Livermore National Laboratory under Contract DE-AC52-07NA27344 and was supported by the Lawrence Fellowship through LLNL-LDRD Program under Project No. 19-ERD-038.


\bibliographystyle{apsrev}

\begin{thebibliography}{55}
	\expandafter\ifx\csname natexlab\endcsname\relax\def\natexlab#1{#1}\fi
	\expandafter\ifx\csname bibnamefont\endcsname\relax
	\def\bibnamefont#1{#1}\fi
	\expandafter\ifx\csname bibfnamefont\endcsname\relax
	\def\bibfnamefont#1{#1}\fi
	\expandafter\ifx\csname citenamefont\endcsname\relax
	\def\citenamefont#1{#1}\fi
	\expandafter\ifx\csname url\endcsname\relax
	\def\url#1{\texttt{#1}}\fi
	\expandafter\ifx\csname urlprefix\endcsname\relax\def\urlprefix{URL }\fi
	\providecommand{\bibinfo}[2]{#2}
	\providecommand{\eprint}[2][]{\url{#2}}
	
	\bibitem[{\citenamefont{Rohrlich}(2000)}]{Rohrlich2000self}
	\bibinfo{author}{\bibfnamefont{F.}~\bibnamefont{Rohrlich}},
	\bibinfo{journal}{Am. J. Phys.} \textbf{\bibinfo{volume}{68}},
	\bibinfo{pages}{1109} (\bibinfo{year}{2000}).
	
	\bibitem[{\citenamefont{McDonald}(2018)}]{Mcdonald2018history}
	\bibinfo{author}{\bibfnamefont{K.~T.} \bibnamefont{McDonald}},
	\emph{\bibinfo{title}{On the history of the radiation reaction}}
	(\bibinfo{year}{2018}).
	
	\bibitem[{\citenamefont{Cole et~al.}(2018)\citenamefont{Cole, Behm, Gerstmayr,
			Blackburn, Wood, Baird, Duff, Harvey, Ilderton, Joglekar
			et~al.}}]{Cole2018experimental}
	\bibinfo{author}{\bibfnamefont{J.~M.} \bibnamefont{Cole}},
	\bibinfo{author}{\bibfnamefont{K.~T.} \bibnamefont{Behm}},
	\bibinfo{author}{\bibfnamefont{E.}~\bibnamefont{Gerstmayr}},
	\bibinfo{author}{\bibfnamefont{T.~G.} \bibnamefont{Blackburn}},
	\bibinfo{author}{\bibfnamefont{J.~C.} \bibnamefont{Wood}},
	\bibinfo{author}{\bibfnamefont{C.~D.} \bibnamefont{Baird}},
	\bibinfo{author}{\bibfnamefont{M.~J.} \bibnamefont{Duff}},
	\bibinfo{author}{\bibfnamefont{C.}~\bibnamefont{Harvey}},
	\bibinfo{author}{\bibfnamefont{A.}~\bibnamefont{Ilderton}},
	\bibinfo{author}{\bibfnamefont{A.~S.} \bibnamefont{Joglekar}},
	\bibnamefont{et~al.}, \bibinfo{journal}{Phys. Rev. X}
	\textbf{\bibinfo{volume}{8}}, \bibinfo{pages}{011020} (\bibinfo{year}{2018}).
	
	\bibitem[{\citenamefont{Poder et~al.}(2018)\citenamefont{Poder, Tamburini,
			Sarri, Di~Piazza, Kuschel, Baird, Behm, Bohlen, Cole, Corvan
			et~al.}}]{Poder2018experimental}
	\bibinfo{author}{\bibfnamefont{K.}~\bibnamefont{Poder}},
	\bibinfo{author}{\bibfnamefont{M.}~\bibnamefont{Tamburini}},
	\bibinfo{author}{\bibfnamefont{G.}~\bibnamefont{Sarri}},
	\bibinfo{author}{\bibfnamefont{A.}~\bibnamefont{Di~Piazza}},
	\bibinfo{author}{\bibfnamefont{S.}~\bibnamefont{Kuschel}},
	\bibinfo{author}{\bibfnamefont{C.~D.} \bibnamefont{Baird}},
	\bibinfo{author}{\bibfnamefont{K.}~\bibnamefont{Behm}},
	\bibinfo{author}{\bibfnamefont{S.}~\bibnamefont{Bohlen}},
	\bibinfo{author}{\bibfnamefont{J.~M.} \bibnamefont{Cole}},
	\bibinfo{author}{\bibfnamefont{D.~J.} \bibnamefont{Corvan}},
	\bibnamefont{et~al.}, \bibinfo{journal}{Phys. Rev. X}
	\textbf{\bibinfo{volume}{8}}, \bibinfo{pages}{031004} (\bibinfo{year}{2018}).
	
	\bibitem[{\citenamefont{Wistisen et~al.}(2018)\citenamefont{Wistisen,
			Di~Piazza, Knudsen, and Uggerh{\o}j}}]{Wistisen2018experimental}
	\bibinfo{author}{\bibfnamefont{T.~N.} \bibnamefont{Wistisen}},
	\bibinfo{author}{\bibfnamefont{A.}~\bibnamefont{Di~Piazza}},
	\bibinfo{author}{\bibfnamefont{H.~V.} \bibnamefont{Knudsen}},
	\bibnamefont{and} \bibinfo{author}{\bibfnamefont{U.~I.}
		\bibnamefont{Uggerh{\o}j}}, \bibinfo{journal}{Nat. Commun.}
	\textbf{\bibinfo{volume}{9}}, \bibinfo{pages}{795} (\bibinfo{year}{2018}).
	
	\bibitem[{\citenamefont{Neitz and Di~Piazza}(2013)}]{Neitz13}
	\bibinfo{author}{\bibfnamefont{N.}~\bibnamefont{Neitz}} \bibnamefont{and}
	\bibinfo{author}{\bibfnamefont{A.}~\bibnamefont{Di~Piazza}},
	\bibinfo{journal}{Phys. Rev. Lett.} \textbf{\bibinfo{volume}{111}},
	\bibinfo{pages}{054802} (\bibinfo{year}{2013}).
	
	\bibitem[{\citenamefont{Blackburn et~al.}(2014)\citenamefont{Blackburn,
			Ridgers, Kirk, and Bell}}]{Blackburn14}
	\bibinfo{author}{\bibfnamefont{T.~G.} \bibnamefont{Blackburn}},
	\bibinfo{author}{\bibfnamefont{C.~P.} \bibnamefont{Ridgers}},
	\bibinfo{author}{\bibfnamefont{J.~G.} \bibnamefont{Kirk}}, \bibnamefont{and}
	\bibinfo{author}{\bibfnamefont{A.~R.} \bibnamefont{Bell}},
	\bibinfo{journal}{Phys. Rev. Lett.} \textbf{\bibinfo{volume}{112}},
	\bibinfo{pages}{015001} (\bibinfo{year}{2014}).
	
	\bibitem[{\citenamefont{Dinu et~al.}(2016)\citenamefont{Dinu, Harvey, Ilderton,
			Marklund, and Torgrimsson}}]{Dinu16}
	\bibinfo{author}{\bibfnamefont{V.}~\bibnamefont{Dinu}},
	\bibinfo{author}{\bibfnamefont{C.}~\bibnamefont{Harvey}},
	\bibinfo{author}{\bibfnamefont{A.}~\bibnamefont{Ilderton}},
	\bibinfo{author}{\bibfnamefont{M.}~\bibnamefont{Marklund}}, \bibnamefont{and}
	\bibinfo{author}{\bibfnamefont{G.}~\bibnamefont{Torgrimsson}},
	\bibinfo{journal}{Phys. Rev. Lett.} \textbf{\bibinfo{volume}{116}},
	\bibinfo{pages}{044801} (\bibinfo{year}{2016}).
	
	\bibitem[{\citenamefont{Kumar et~al.}(2013)\citenamefont{Kumar, Hatsagortsyan,
			and Keitel}}]{Kumar13}
	\bibinfo{author}{\bibfnamefont{N.}~\bibnamefont{Kumar}},
	\bibinfo{author}{\bibfnamefont{K.~Z.} \bibnamefont{Hatsagortsyan}},
	\bibnamefont{and} \bibinfo{author}{\bibfnamefont{C.~H.}
		\bibnamefont{Keitel}}, \bibinfo{journal}{Phys. Rev. Lett.}
	\textbf{\bibinfo{volume}{111}}, \bibinfo{pages}{105001}
	(\bibinfo{year}{2013}).
	
	\bibitem[{\citenamefont{Ji et~al.}(2014)\citenamefont{Ji, Pukhov, Kostyukov,
			Shen, and Akli}}]{Ji14}
	\bibinfo{author}{\bibfnamefont{L.~L.} \bibnamefont{Ji}},
	\bibinfo{author}{\bibfnamefont{A.}~\bibnamefont{Pukhov}},
	\bibinfo{author}{\bibfnamefont{I.~Y.} \bibnamefont{Kostyukov}},
	\bibinfo{author}{\bibfnamefont{B.~F.} \bibnamefont{Shen}}, \bibnamefont{and}
	\bibinfo{author}{\bibfnamefont{K.}~\bibnamefont{Akli}},
	\bibinfo{journal}{Phys. Rev. Lett.} \textbf{\bibinfo{volume}{112}},
	\bibinfo{pages}{145003} (\bibinfo{year}{2014}).
	
	\bibitem[{\citenamefont{Liseykina et~al.}(2016)\citenamefont{Liseykina,
			Popruzhenko, and Macchi}}]{Liseykina2016inverse}
	\bibinfo{author}{\bibfnamefont{T.~V.} \bibnamefont{Liseykina}},
	\bibinfo{author}{\bibfnamefont{S.~V.} \bibnamefont{Popruzhenko}},
	\bibnamefont{and} \bibinfo{author}{\bibfnamefont{A.}~\bibnamefont{Macchi}},
	\bibinfo{journal}{New J. Phys.} \textbf{\bibinfo{volume}{18}},
	\bibinfo{pages}{072001} (\bibinfo{year}{2016}).
	
	\bibitem[{\citenamefont{Ritus}(1985)}]{Ritus1985}
	\bibinfo{author}{\bibfnamefont{V.~I.} \bibnamefont{Ritus}},
	\bibinfo{journal}{J. Sov. Laser Res.} \textbf{\bibinfo{volume}{6}},
	\bibinfo{pages}{497} (\bibinfo{year}{1985}).
	
	\bibitem[{\citenamefont{Bula et~al.}(1996)\citenamefont{Bula, McDonald, Prebys,
			Bamber, Boege, Kotseroglou, Melissinos, Meyerhofer, Ragg, Burke
			et~al.}}]{Bula1996observation}
	\bibinfo{author}{\bibfnamefont{C.}~\bibnamefont{Bula}},
	\bibinfo{author}{\bibfnamefont{K.~T.} \bibnamefont{McDonald}},
	\bibinfo{author}{\bibfnamefont{E.~J.} \bibnamefont{Prebys}},
	\bibinfo{author}{\bibfnamefont{C.}~\bibnamefont{Bamber}},
	\bibinfo{author}{\bibfnamefont{S.}~\bibnamefont{Boege}},
	\bibinfo{author}{\bibfnamefont{T.}~\bibnamefont{Kotseroglou}},
	\bibinfo{author}{\bibfnamefont{A.~C.} \bibnamefont{Melissinos}},
	\bibinfo{author}{\bibfnamefont{D.~D.} \bibnamefont{Meyerhofer}},
	\bibinfo{author}{\bibfnamefont{W.}~\bibnamefont{Ragg}},
	\bibinfo{author}{\bibfnamefont{D.~L.} \bibnamefont{Burke}},
	\bibnamefont{et~al.}, \bibinfo{journal}{Phys. Rev. Lett.}
	\textbf{\bibinfo{volume}{76}}, \bibinfo{pages}{3116} (\bibinfo{year}{1996}).
	
	\bibitem[{\citenamefont{Hartemann and Kerman}(1996)}]{Hartemann1996classical}
	\bibinfo{author}{\bibfnamefont{F.~V.} \bibnamefont{Hartemann}}
	\bibnamefont{and} \bibinfo{author}{\bibfnamefont{A.~K.}
		\bibnamefont{Kerman}}, \bibinfo{journal}{Phys. Rev. Lett.}
	\textbf{\bibinfo{volume}{76}}, \bibinfo{pages}{624} (\bibinfo{year}{1996}).
	
	\bibitem[{\citenamefont{Keitel et~al.}(1998)\citenamefont{Keitel, Szymanowski,
			Knight, and Maquet}}]{Keitel1998radiative}
	\bibinfo{author}{\bibfnamefont{C.~H.} \bibnamefont{Keitel}},
	\bibinfo{author}{\bibfnamefont{C.}~\bibnamefont{Szymanowski}},
	\bibinfo{author}{\bibfnamefont{P.~L.} \bibnamefont{Knight}},
	\bibnamefont{and} \bibinfo{author}{\bibfnamefont{A.}~\bibnamefont{Maquet}},
	\bibinfo{journal}{J. Phys. B-At. Mol. Opt.} \textbf{\bibinfo{volume}{31}},
	\bibinfo{pages}{L75} (\bibinfo{year}{1998}).
	
	\bibitem[{\citenamefont{Di~Piazza et~al.}(2009)\citenamefont{Di~Piazza,
			Hatsagortsyan, and Keitel}}]{Piazza2009strong}
	\bibinfo{author}{\bibfnamefont{A.}~\bibnamefont{Di~Piazza}},
	\bibinfo{author}{\bibfnamefont{K.~Z.} \bibnamefont{Hatsagortsyan}},
	\bibnamefont{and} \bibinfo{author}{\bibfnamefont{C.~H.}
		\bibnamefont{Keitel}}, \bibinfo{journal}{Phys. Rev. Lett.}
	\textbf{\bibinfo{volume}{102}}, \bibinfo{pages}{254802}
	(\bibinfo{year}{2009}).
	
	\bibitem[{\citenamefont{Di~Piazza et~al.}(2010)\citenamefont{Di~Piazza,
			Hatsagortsyan, and Keitel}}]{Piazza10}
	\bibinfo{author}{\bibfnamefont{A.}~\bibnamefont{Di~Piazza}},
	\bibinfo{author}{\bibfnamefont{K.~Z.} \bibnamefont{Hatsagortsyan}},
	\bibnamefont{and} \bibinfo{author}{\bibfnamefont{C.~H.}
		\bibnamefont{Keitel}}, \bibinfo{journal}{Phys. Rev. Lett.}
	\textbf{\bibinfo{volume}{105}}, \bibinfo{pages}{220403}
	(\bibinfo{year}{2010}).
	
	\bibitem[{\citenamefont{Kravets et~al.}(2013)\citenamefont{Kravets, Noble, and
			Jaroszynski}}]{Kravets13}
	\bibinfo{author}{\bibfnamefont{Y.}~\bibnamefont{Kravets}},
	\bibinfo{author}{\bibfnamefont{A.}~\bibnamefont{Noble}}, \bibnamefont{and}
	\bibinfo{author}{\bibfnamefont{D.}~\bibnamefont{Jaroszynski}},
	\bibinfo{journal}{Phys. Rev. E} \textbf{\bibinfo{volume}{88}},
	\bibinfo{pages}{011201} (\bibinfo{year}{2013}).
	
	\bibitem[{\citenamefont{Li et~al.}(2014)\citenamefont{Li, Hatsagortsyan, and
			Keitel}}]{Li14}
	\bibinfo{author}{\bibfnamefont{J.-X.} \bibnamefont{Li}},
	\bibinfo{author}{\bibfnamefont{K.~Z.} \bibnamefont{Hatsagortsyan}},
	\bibnamefont{and} \bibinfo{author}{\bibfnamefont{C.~H.}
		\bibnamefont{Keitel}}, \bibinfo{journal}{Phys. Rev. Lett.}
	\textbf{\bibinfo{volume}{113}}, \bibinfo{pages}{044801}
	(\bibinfo{year}{2014}).
	
	\bibitem[{\citenamefont{Fulton and Rohrlich}(1960)}]{Fulton1960classical}
	\bibinfo{author}{\bibfnamefont{T.}~\bibnamefont{Fulton}} \bibnamefont{and}
	\bibinfo{author}{\bibfnamefont{F.}~\bibnamefont{Rohrlich}},
	\bibinfo{journal}{Ann. Phys.} \textbf{\bibinfo{volume}{9}},
	\bibinfo{pages}{499} (\bibinfo{year}{1960}).
	
	\bibitem[{\citenamefont{Rohrlich}(1961{\natexlab{a}})}]{Rohrlich1961definition}
	\bibinfo{author}{\bibfnamefont{F.}~\bibnamefont{Rohrlich}},
	\bibinfo{journal}{Nuovo Cim.} \textbf{\bibinfo{volume}{21}},
	\bibinfo{pages}{811} (\bibinfo{year}{1961}{\natexlab{a}}).
	
	\bibitem[{\citenamefont{Boulware}(1980)}]{Boulware1980radiation}
	\bibinfo{author}{\bibfnamefont{D.~G.} \bibnamefont{Boulware}},
	\bibinfo{journal}{Ann. Phys.} \textbf{\bibinfo{volume}{124}},
	\bibinfo{pages}{169} (\bibinfo{year}{1980}).
	
	\bibitem[{\citenamefont{Schild}(1960)}]{Schild1960radiation}
	\bibinfo{author}{\bibfnamefont{A.}~\bibnamefont{Schild}}, \bibinfo{journal}{J.
		Math. Anal. Appl.} \textbf{\bibinfo{volume}{1}}, \bibinfo{pages}{1271}
	(\bibinfo{year}{1960}).
	
	\bibitem[{\citenamefont{Lorentz}(1892)}]{Lorentz1892theorie}
	\bibinfo{author}{\bibfnamefont{H.~A.} \bibnamefont{Lorentz}},
	\emph{\bibinfo{title}{La th{\'e}orie {\'e}lectromagn{\'e}tique de {M}axwell
			et son application aux corps mouvants}} (\bibinfo{publisher}{E J Brill},
	\bibinfo{year}{1892}).
	
	\bibitem[{\citenamefont{Abraham}(1905)}]{Abraham1905theorie}
	\bibinfo{author}{\bibfnamefont{M.}~\bibnamefont{Abraham}},
	\emph{\bibinfo{title}{Theorie der Elektrizit{\"a}t Zweiter Band:
			Elektromagnetische Theorie der Strahlung}} (\bibinfo{publisher}{B G Teubner,
		Leipzig}, \bibinfo{year}{1905}).
	
	\bibitem[{\citenamefont{Dirac}(1938)}]{Dirac1938classical}
	\bibinfo{author}{\bibfnamefont{P.~A.~M.} \bibnamefont{Dirac}},
	\bibinfo{journal}{Proc. R. Soc. Lond. Ser. A} \textbf{\bibinfo{volume}{167}},
	\bibinfo{pages}{148} (\bibinfo{year}{1938}).
	
	\bibitem[{\citenamefont{Coleman}(1961)}]{coleman1961classical}
	\bibinfo{author}{\bibfnamefont{S.}~\bibnamefont{Coleman}}, \bibinfo{type}{Tech.
		Rep.}, \bibinfo{institution}{{RAND} Corp. Santa Monica Calif.}
	(\bibinfo{year}{1961}).
	
	\bibitem[{\citenamefont{Krivitski\u{i} and Tsytovich}(1991)}]{Krivitskii91}
	\bibinfo{author}{\bibfnamefont{V.~S.} \bibnamefont{Krivitski\u{i}}}
	\bibnamefont{and} \bibinfo{author}{\bibfnamefont{V.~N.}
		\bibnamefont{Tsytovich}}, \bibinfo{journal}{Sov. Phys. Usp.}
	\textbf{\bibinfo{volume}{34}}, \bibinfo{pages}{250} (\bibinfo{year}{1991}).
	
	\bibitem[{\citenamefont{Ilderton and
			Torgrimsson}(2013{\natexlab{a}})}]{Ilderton13PLB}
	\bibinfo{author}{\bibfnamefont{A.}~\bibnamefont{Ilderton}} \bibnamefont{and}
	\bibinfo{author}{\bibfnamefont{G.}~\bibnamefont{Torgrimsson}},
	\bibinfo{journal}{Phys. Lett. B} \textbf{\bibinfo{volume}{725}},
	\bibinfo{pages}{481} (\bibinfo{year}{2013}{\natexlab{a}}).
	
	\bibitem[{\citenamefont{Ilderton and
			Torgrimsson}(2013{\natexlab{b}})}]{Ilderton13PRD}
	\bibinfo{author}{\bibfnamefont{A.}~\bibnamefont{Ilderton}} \bibnamefont{and}
	\bibinfo{author}{\bibfnamefont{G.}~\bibnamefont{Torgrimsson}},
	\bibinfo{journal}{Phys. Rev. D} \textbf{\bibinfo{volume}{88}},
	\bibinfo{pages}{025021} (\bibinfo{year}{2013}{\natexlab{b}}).
	
	\bibitem[{\citenamefont{Eriksen and Gr{\o}n}(2000)}]{Eriksen00}
	\bibinfo{author}{\bibfnamefont{E.}~\bibnamefont{Eriksen}} \bibnamefont{and}
	\bibinfo{author}{\bibfnamefont{{\O}.}~\bibnamefont{Gr{\o}n}},
	\bibinfo{journal}{Ann. Phys.} \textbf{\bibinfo{volume}{286}},
	\bibinfo{pages}{373} (\bibinfo{year}{2000}).
	
	\bibitem[{\citenamefont{Schott}(1912)}]{schott1912electromagnetic}
	\bibinfo{author}{\bibfnamefont{G.~A.} \bibnamefont{Schott}},
	\emph{\bibinfo{title}{Electromagnetic Radiation and the Mechanical Reactions
			Arising from It}} (\bibinfo{publisher}{Cambridge University Press},
	\bibinfo{year}{1912}).
	
	\bibitem[{\citenamefont{Teitelboim}(1970)}]{Teitelboim70}
	\bibinfo{author}{\bibfnamefont{C.}~\bibnamefont{Teitelboim}},
	\bibinfo{journal}{Phys. Rev. D} \textbf{\bibinfo{volume}{1}},
	\bibinfo{pages}{1572} (\bibinfo{year}{1970}).
	
	\bibitem[{\citenamefont{Bondi and Gold}(1955)}]{Bondi1955field}
	\bibinfo{author}{\bibfnamefont{H.}~\bibnamefont{Bondi}} \bibnamefont{and}
	\bibinfo{author}{\bibfnamefont{T.}~\bibnamefont{Gold}},
	\bibinfo{journal}{Proc. R. Soc. Lond. Ser. A} \textbf{\bibinfo{volume}{229}},
	\bibinfo{pages}{416} (\bibinfo{year}{1955}).
	
	\bibitem[{\citenamefont{Gr{\o}n}(2011)}]{Gron2011significance}
	\bibinfo{author}{\bibfnamefont{{\O}.}~\bibnamefont{Gr{\o}n}},
	\bibinfo{journal}{Am. J. Phys.} \textbf{\bibinfo{volume}{79}},
	\bibinfo{pages}{115} (\bibinfo{year}{2011}).
	
	\bibitem[{\citenamefont{Ng}(1993)}]{Ng93}
	\bibinfo{author}{\bibfnamefont{C.-S.} \bibnamefont{Ng}},
	\bibinfo{journal}{Phys. Rev. E} \textbf{\bibinfo{volume}{47}},
	\bibinfo{pages}{2038} (\bibinfo{year}{1993}).
	
	\bibitem[{\citenamefont{Eriksen and
			Gr{\o}n}(2002)}]{Eriksen2002electrodynamics}
	\bibinfo{author}{\bibfnamefont{E.}~\bibnamefont{Eriksen}} \bibnamefont{and}
	\bibinfo{author}{\bibfnamefont{{\O}.}~\bibnamefont{Gr{\o}n}},
	\bibinfo{journal}{Ann. Phys.} \textbf{\bibinfo{volume}{297}},
	\bibinfo{pages}{243} (\bibinfo{year}{2002}).
	
	\bibitem[{\citenamefont{Yaghjian}(2006)}]{Yaghjian2010relativistic}
	\bibinfo{author}{\bibfnamefont{A.}~\bibnamefont{Yaghjian}},
	\emph{\bibinfo{title}{Relativistic dynamics of a charged sphere: updating the
			{L}orentz-{A}braham model}}, vol.~\bibinfo{volume}{11}
	(\bibinfo{publisher}{Springer-Verlag New York}, \bibinfo{year}{2006}).
	
	\bibitem[{\citenamefont{Rohrlich}(1997)}]{Rohrlich1997dynamics}
	\bibinfo{author}{\bibfnamefont{F.}~\bibnamefont{Rohrlich}},
	\bibinfo{journal}{Am. J. Phys.} \textbf{\bibinfo{volume}{65}},
	\bibinfo{pages}{1051} (\bibinfo{year}{1997}).
	
	\bibitem[{\citenamefont{Smorenburg et~al.}(2014)\citenamefont{Smorenburg, Kamp,
			and Luiten}}]{Smorenburg2014classical}
	\bibinfo{author}{\bibfnamefont{P.~W.} \bibnamefont{Smorenburg}},
	\bibinfo{author}{\bibfnamefont{L.~P.~J.} \bibnamefont{Kamp}},
	\bibnamefont{and} \bibinfo{author}{\bibfnamefont{O.~J.}
		\bibnamefont{Luiten}}, \bibinfo{journal}{Eur. Phys. J. H}
	\textbf{\bibinfo{volume}{39}}, \bibinfo{pages}{283} (\bibinfo{year}{2014}).
	
	\bibitem[{\citenamefont{Bhabha and Corben}(1941)}]{Bhabha1941general}
	\bibinfo{author}{\bibfnamefont{H.~J.} \bibnamefont{Bhabha}} \bibnamefont{and}
	\bibinfo{author}{\bibfnamefont{H.~C.} \bibnamefont{Corben}},
	\bibinfo{journal}{Proc. R. Soc. Lond. Ser. A} \textbf{\bibinfo{volume}{178}},
	\bibinfo{pages}{273} (\bibinfo{year}{1941}).
	
	\bibitem[{\citenamefont{Rowe}(1975)}]{Rowe75}
	\bibinfo{author}{\bibfnamefont{E.~G.~P.} \bibnamefont{Rowe}},
	\bibinfo{journal}{Phys. Rev. D} \textbf{\bibinfo{volume}{12}},
	\bibinfo{pages}{1576} (\bibinfo{year}{1975}).
	
	\bibitem[{\citenamefont{Honig and Szamosi}(1983)}]{Honig1983general}
	\bibinfo{author}{\bibfnamefont{E.}~\bibnamefont{Honig}} \bibnamefont{and}
	\bibinfo{author}{\bibfnamefont{G.}~\bibnamefont{Szamosi}},
	\bibinfo{journal}{Phys. Lett. A} \textbf{\bibinfo{volume}{93}},
	\bibinfo{pages}{319} (\bibinfo{year}{1983}).
	
	\bibitem[{\citenamefont{Lozada}(1989)}]{Lozada1989general}
	\bibinfo{author}{\bibfnamefont{A.}~\bibnamefont{Lozada}}, \bibinfo{journal}{J.
		Math. Phys.} \textbf{\bibinfo{volume}{30}}, \bibinfo{pages}{1713}
	(\bibinfo{year}{1989}).
	
	\bibitem[{\citenamefont{Aguirregabiria
			et~al.}(2006)\citenamefont{Aguirregabiria, Llosa, and
			Molina}}]{Aguirregabiria06}
	\bibinfo{author}{\bibfnamefont{J.~M.} \bibnamefont{Aguirregabiria}},
	\bibinfo{author}{\bibfnamefont{J.}~\bibnamefont{Llosa}}, \bibnamefont{and}
	\bibinfo{author}{\bibfnamefont{A.}~\bibnamefont{Molina}},
	\bibinfo{journal}{Phys. Rev. D} \textbf{\bibinfo{volume}{73}},
	\bibinfo{pages}{125015} (\bibinfo{year}{2006}).
	
	\bibitem[{\citenamefont{Medina}(2006)}]{Medina2006radiation}
	\bibinfo{author}{\bibfnamefont{R.}~\bibnamefont{Medina}}, \bibinfo{journal}{J.
		Phys. A: Math. Gen.} \textbf{\bibinfo{volume}{39}}, \bibinfo{pages}{3801}
	(\bibinfo{year}{2006}).
	
	\bibitem[{\citenamefont{Moniz and Sharp}(1977)}]{Moniz1977radiation}
	\bibinfo{author}{\bibfnamefont{E.~J.} \bibnamefont{Moniz}} \bibnamefont{and}
	\bibinfo{author}{\bibfnamefont{D.~H.} \bibnamefont{Sharp}},
	\bibinfo{journal}{Phys. Rev. D} \textbf{\bibinfo{volume}{15}},
	\bibinfo{pages}{2850} (\bibinfo{year}{1977}).
	
	\bibitem[{\citenamefont{Johnson and Hu}(2002)}]{Johnson2002stochastic}
	\bibinfo{author}{\bibfnamefont{P.~R.} \bibnamefont{Johnson}} \bibnamefont{and}
	\bibinfo{author}{\bibfnamefont{B.-L.} \bibnamefont{Hu}},
	\bibinfo{journal}{Phys. Rev. D} \textbf{\bibinfo{volume}{65}},
	\bibinfo{pages}{065015} (\bibinfo{year}{2002}).
	
	\bibitem[{\citenamefont{Eliezer}(1948)}]{Eliezer1948classical}
	\bibinfo{author}{\bibfnamefont{C.~J.} \bibnamefont{Eliezer}},
	\bibinfo{journal}{Proc. R. Soc. Lond. Ser. A} \textbf{\bibinfo{volume}{194}},
	\bibinfo{pages}{543} (\bibinfo{year}{1948}).
	
	\bibitem[{\citenamefont{Landau and Lifshitz}(1971)}]{Landau1971classical}
	\bibinfo{author}{\bibfnamefont{L.~D.} \bibnamefont{Landau}} \bibnamefont{and}
	\bibinfo{author}{\bibfnamefont{E.~M.} \bibnamefont{Lifshitz}},
	\emph{\bibinfo{title}{The classical theory of fields}}
	(\bibinfo{publisher}{Pergamon}, \bibinfo{year}{1971}).
	
	\bibitem[{\citenamefont{Spohn}(2000)}]{Spohn2000critical}
	\bibinfo{author}{\bibfnamefont{H.}~\bibnamefont{Spohn}},
	\bibinfo{journal}{Europhys. Lett.} \textbf{\bibinfo{volume}{50}},
	\bibinfo{pages}{287} (\bibinfo{year}{2000}).
	
	\bibitem[{\citenamefont{Mo and Papas}(1971)}]{Mo71}
	\bibinfo{author}{\bibfnamefont{T.~C.} \bibnamefont{Mo}} \bibnamefont{and}
	\bibinfo{author}{\bibfnamefont{C.~H.} \bibnamefont{Papas}},
	\bibinfo{journal}{Phys. Rev. D} \textbf{\bibinfo{volume}{4}},
	\bibinfo{pages}{3566} (\bibinfo{year}{1971}).
	
	\bibitem[{\citenamefont{Ford and O'Connell}(1991)}]{Ford1991radiation}
	\bibinfo{author}{\bibfnamefont{G.~W.} \bibnamefont{Ford}} \bibnamefont{and}
	\bibinfo{author}{\bibfnamefont{R.~F.} \bibnamefont{O'Connell}},
	\bibinfo{journal}{Phys. Lett. A} \textbf{\bibinfo{volume}{157}},
	\bibinfo{pages}{217} (\bibinfo{year}{1991}).
	
	\bibitem[{\citenamefont{Rohrlich}(2002)}]{Rohrlich2002dynamics}
	\bibinfo{author}{\bibfnamefont{F.}~\bibnamefont{Rohrlich}},
	\bibinfo{journal}{Phys. Lett. A} \textbf{\bibinfo{volume}{303}},
	\bibinfo{pages}{307} (\bibinfo{year}{2002}).
	
	\bibitem[{\citenamefont{Rohrlich}(1961{\natexlab{b}})}]{Rohrlich1961equations}
	\bibinfo{author}{\bibfnamefont{F.}~\bibnamefont{Rohrlich}},
	\bibinfo{journal}{Ann. Phys.} \textbf{\bibinfo{volume}{13}},
	\bibinfo{pages}{93} (\bibinfo{year}{1961}{\natexlab{b}}).
	
\end{thebibliography}







\end{document}